\documentclass[11pt,twoside,a4paper,reqno]{amsart}
\usepackage[cp1251]{inputenc}
\usepackage[greek,russian,english]{babel}
\usepackage{amssymb,graphicx,color}
\usepackage{mathrsfs,dsfont}
\usepackage{mathpazo}

\usepackage[pdfauthor={Brezhnev}, colorlinks=true, linktocpage=false,
						pdfwindowui=false, pdfmenubar=false, citecolor=cyan,
						linkcolor=blue, hyperfootnotes=true, pdfstartview=FitH,
						pdfkeywords={quantum foundations, non-axiomaticity,
						bases of observables, orthogonality, scalar product,
						Hilbert space, Pythagoras theorem, Hilbert 6th problem},
						bookmarksopen=false, bookmarks=false]{hyperref}

\usepackage{CutPage}
\textcenter{155mm}{219mm}
\usepackage{YURA}
\usepackage{Foots}

\theoremstyle{remark}%
\newtheorem{comm}{%
\indent\blue \textup{{R\,[3]e\,[3]m\,[3]a\,[3]r\,[3]k}}~~\ignorespaces}
\newenvironment{comment}%
{\par\smaller[1]\begin{comm}}{\end{comm}}

\newcommand{\brak}[3][|]{\hbox to0ex{$\langle$}\mkern2mu\langle#2%
#1\mkern1mu#3\hbox to0ex{$\rangle$}\mkern2mu\rangle}
\newcommand{\GOTO}[2][2]{\mathrel{
	\Over[1]{#2}{\REPEAT{\mathchar"439}{#1} \mathchar"044B}}}

\def\+{\mathbin{\hat+}}
\def\T{{\raise0.15ex\hbox%
	{$\Over[0.4]{\rule{2.8pt}{0.45pt}}{\rule{0.45pt}{2pt}}$}}}

\newcommand{\DOWN}[1][2]{\rotatebox[origin=cc]{-90}{$\goto[#1]$}}

\def\tgreek#1{\textgreek{\texttt{#1}}}
\def\state#1{{\,\underline{\!#1\!}\,}{}}
\def\ket#1{|\bo#1\rangle}
\def\bra#1{\langle\bo#1|}
\def\smallket#1{{\sss|\bo#1\rangle}}

\def\fr{\mathtt{f\!}}

\def\StatL{\textup{\texttt{Stat\-Length}}}
\def\boe{\mathbf{e}}
\def\statePsi{\state\Psi}
\def\ketPsi{\ket\Psi}
\def\ketPhi{\ket\Phi}

\def\hA{\oper[17]{\scr A}}
\def\hB{\oper[6] {\scr B}}

\newcommand{\Jref}{%
	\href{https://doi.org/10.1016/j.geomphys.2023.104779}
	{\cyan\emph{J.~Geom.~Phys.} (2023) \textbf{187}, 104779}}

\author[\hfill\textsf{Yu.~Brezhnev}\hfill\llap{\smaller[2]\Jref}]%
{Yurii~V.~Brezhnev$\red^{\bo*}$}

\title[\rlap{\smaller[2]\Jref}\hfill
\textsl{W\lowercase{hy \uppercase{H}ilbert space?\hfill}}]%
{Why and whence the Hilbert space in quantum theory?}

\address{Department of Quantum Field Theory,
Tomsk State University, Russia}
\email{\Courier{brezhnev@phys.tsu.ru}}

\def\datename{\hspace{-1.5em}\relax}
\date{\setlength{\arraycolsep}{0em}%
$\begin{array}[b]{rll}
&\text{\emph{Key words}: }&
\makebox[0em][l]{quantum foundations, non-axiomaticity,
bases of observables, orthogonality, scalar product,}\\
&&\makebox[0em][l]{topology, Hilbert space,
Pythagoras theorem, hierarchy of math/physical languages}\\[1ex]
\red^{\bo*}&\makebox[0em][l]{\textsf{Department of Quantum Field
Theory, Tomsk State University, Russia}}\\
&\makebox[0em][l]{\Courier{brezhnev@phys.tsu.ru}}
\end{array}$
\hfill\smaller[2]1 March 2023\![5]}

\begin{document}
\ \vspace{-6ex}

{\smaller[2](\Russian русская версия по email-запросу)}\hfill
{\href{https://arxiv.org/abs/2110.05932}%
{\smaller[2]\cyan\Courier{https:/\!\!\!/arxiv.org/abs/2110.05932}}}
\\[-3ex]

\hfill{\smaller[1]
\href{https://doi.org/10.1016/j.geomphys.2023.104779}
{\smaller\cyan\emph{Journ.~Geom.~Phys.} (2023) \textbf{187}(May),
104779~(15)}}\\

\begin{abstract}
\vspace{-2ex}
We explain why and how the Hilbert space comes about in quantum
theory. The axiomatic structures of vector space, of scalar product,
of orthogonality, and of the linear functional are derivable from
the statistical description of quantum micro-events and from
Hilbertian sum of squares $|\frak a_1|^2+|\frak a_2|^2+\cdots$. The
latter leads (non"=axiomatically) to the standard writing of the Born
formula $\fr=|\langle\psi|\varphi\rangle|^2$. As a corollary, the
status of Pythagorean theorem, the concept of a length, and the 6"~th
Hilbert problem undergo a quantum `revision'. An issue of deriving
the norm topology may no have a short-length solution (too many
abstract math"=axioms) but is likely solvable in the affirmative; the
problem is reformulated as a
mathematical one.\\[-4ex]
\end{abstract}

\maketitle
\thispagestyle{empty}

{\def\footnotemark{\relax}
\smaller[1]\tableofcontents}
\newpage

\vbox{\flushright\smaller[1.5]%
\textsl{I would like to make a confession \ldots: I do not\\ believe
absolutely in Hilbert space any more}\\[1.5ex]
\textsl{I, for one, do not even believe, that the right formal\\
frame for quantum mechanics is already found}\\[1.5ex]
\textsc{--- J.~von~Neumann} to \textsc{G.~Birkhoff} (1935--36)}

\section{Introduction}\label{intro}

The machinery of quantum theory (\qt) starts with the Hilbert space
and self"=adjoint operators acting on it. The question of whence
these very mathematical constructs come as axioms \cite{fuchs2,
wightman} has been actively discussing in the literature hitherto
\cite{laloe, engesser, cassineli, muller, dAriano+}; because of its
profound importance for the theory itself \cite{peres, hren}.

At the same time, leaving aside the quantization of models, the most
part of the theory is self"=sufficiently described only by a linear
vector space (\lvs), \ie, turns into an empirical theorem \cite{br1}
known as the principle of linear superposition of quantum states
over the field of complex numbers~$\bbC$. In doing so, the `quantum
field' $\bbC$ is naturally equipped with the complex"=conjugation
operation $(\msf n+\ri\,\msf m) \mathrel{\Over{*}{\mapsto}}
(\msf n-\ri\,\msf m)$, and for understanding the `primary/"!derivative'
in the quantum-math abstractions""---""vector space, linear
operators, \thelike---the following is necessary. The abstractions
come from the number entities \cite{lakoff}, and even the notion of
the number itself is a nontrivial point in foundations of \qt.
Accordingly, ideology of deducing the \qt"=structures should be
considered `through the lenses' of coordinate representations of
\lvs, inasmuch as initially we have no motives to introduce any
abstracta. For instance, the complex scalar product, along with the
$\bbC$"=numbers themselves, do still leave the issue of its own
nature \cite{accardi+, engesser}, while the real case is seen as
something quite natural. See, however, discussion about `identity'
between the classical Pythagoras' theorem and the quantum Born rule
in \cite{br4}.

Yet a further salient feature lies in the fact that the genesis of
quantum structures cannot in principle be physical""---a point, long
championed by G.~Ludwig \cite[Foreword]{ludwig5}, \cite{ludwig3} and
voiced clearly by other writers (E.~Jaynes, A.~Peres, D.~Mermin; see
references and quotations in ref.~\cite{br1}). By this, one should
understand that the justification in the physics'
tongue---""temporal $t$"=evolution, interaction with environment,
collapses of states, notion of the `state
$|\texttt{before}\rangle$/$|\texttt{after}\rangle$', modelling the
measuring process \cite{busch}, \etc---will always suffer from the
circular logic when attempting to derive mathematical apparatus of
\qt.

The situation is well known in the literature as the `never"=ending'
debate over interpretations \cite{laloe, hren, aaronson+,
silverman}. This has an impact on physicality of argument and even
on the `physical level of rigor' in reasoning; this term should also
be excluded.
\begin{itemize}
\item The search for underpinning of quantum the\-ory---unusual as it may
 seem---is largely \emph{not a question of physics} \cite{ludwig5},
 much less of phenomenology: particles, phenomena, fields and their
 interactions.
\end{itemize}
The similar argument is already present in the literature
\cite[p.~220]{aaronson+}, \cite{hren}. For example, in abstract of
the article \cite{dAriano}, one openly claims that ``mathematization
of the physical theory \ldots\ must contain no physical primitives. In
provocative words: `physics from no physics'\,\,''.

Attempts to justify the formalism of \qt\ \cite{accardi+, dAriano}
had already begun under von~Neumann as a programme called
``continuous geometry'' \cite{neumann}. However, it is now recognized
\cite{kronz} that the programme had not been successful, though it
gave birth to the three mathematically nice theories: algebraic
approach to \qm, quantum logic \cite{engesser}, and rings of
operators (von~Neumann algebras). In this connection we emphasize
that the underpinning cannot constitute the ready"=made mathematics
\cite{br1}, because the mathematical structures, even commonly"=used,
are not delivered from above. The questions `whence/"!why?\@' can
hardly be answered in the sequence \cite{salmon}
\begin{gather}\label{phil-}
\text{\ceils{terminology, definienda} $\rightarrowtail$
\ceils{statement} $\rightarrowtail$ \ceils{proof}}\\
\text{\big(\;\tplus\;\ceils{philosophical/"!physical `support'}\big)}\;.
\notag
\end{gather}

As we will see below, the math"=structures have an origin, and
axiomatizations, so prevalent in \qt\ \cite{engesser, hren, laloe,
ludwig5, peres, silverman}, are not necessary and should ideally be
minimized; an orthodox opinion \cite{accardi+, cassineli, dAriano+,
fuchs2, laloe, muller}. Not only should the mathematical and
physical phrasing not be leaned upon the postulates, but the
language itself should be severely limited \cite{ludwig5} and far
from being free as in the classical theories. The problems with
interpretations of quantum mathematics in terms of observables
\cite{laloe, hren} do not then arise since their language has not
yet in place.

The answer to Neumann's doubts quoted above (and oft mentioned in
the literature \cite{engesser}) lies in the fact that Hilbert's
space \emph{is not} a starting point in quantum elements. For
example, Ludwig has devoted the whole monograph \cite{ludwig3} to
the `quantum' derivation (different from ours) of this structure.
The different views of the Hilbert"=space problem and exhaustive
references along the lines can be found in the works \cite{dAriano+,
hren0, muller}. In the first place, the space is \emph{just a linear
space} (superposition principle), of which the empirical nature
manifests initially as a commutative group with operator
automorphisms \cite[sect.~I.1.2]{kurosh}. These are the numbers, and
focusing on them above is no accident.

Such a (re)formulation, at first glance, might have been seen as a
full abstraction (see \cite[sects.~7--8]{br1} for details), but one
may go further. It is the quantum paradigm and `building' the
quantum mathematics from scratch, rather than a reliance upon the
ready-to-use formal constructs, that provide the most convincing
answer to the question of \emph{why} and \emph{whence} the complex
numbers \cite{cassineli} and the very vector space. The number in
quantum foundations, not only the complex, is not a matter of course
in the context of the accustomed arithmetic \cite{lakoff}. The
subsequent putting the question about observables as of entities and
of their numerical values will result in answer to the question
posed in the title of the work. With that, the observables arise
alongside the state space but not yet as (Hermitian) operators.

The objective of this work is to focus on an end"=point for the
Hilbert-space problem; stated on p.~20 of the previous study
\cite{br2}. The case in point is a derivation of what defines the
Hilbertian \emph{add-ons} over quantum \lvs: scalar product
$\langle\psi | \varphi\rangle$, orthogonality
$\ket{{}\psi}\bot\ket{{}\varphi}$, norm $\|\psi\|$, topology, and
bases of observables. In doing so, we do not assume""---even in a
hidden form---any additional mathematical structures or physical
preconceptions/"!arguments: axioms/"!theorems of probability, quantum
logic, ``assertions about the physical world'', notably the word
`measurement', physical observations, \etc. More precisely, one
deduces an a~priory unknown mathematics that stems from quantumness
of the state"=space and is dictated by quantum paradigm itself. The
follow"=up relationship between quantum Hilbert' space and physics is
discussed in the final section of the work.

\section{Structures on quantum-state space}

The basis for a vector space does always exist \cite{fried, halm},
but the \lvs"=axiomatics, in and of itself, contains neither such a
concept nor a motivation as to why/"!what-for the basis may/"!needs-to
be changed. Similarly, the numeric quantities that are associated
with quantum vectors""---lengths and projections""---do not follow
from anywhere and hence may arise in the \lvs' theory only from the
outside.

\subsection{Bases in quantum theory (why?\@)}

Every \lvs\ possesses the infinitely many bases but \emph{quantum}
space $\bbH$ comes into being \emph{at the outset together with}
these objects \cite{br1}. More than that, it arises through the
special kind bases, termed the $\scr A$"=bases, such that each vector
$\scr A$"=representation
\begin{equation}\label{=}
\ketPsi=\frak a_1\bcdot\ket{\alpha_1}\+
\frak a_2\bcdot \ket{\alpha_2}\+\cdots\quad\in\bbH
\end{equation}
has an associated number characteristic""---$\{\fr_j\}$"=statistics
of quantum $\scr A$"=micro"=events $\state\alpha_j$ (pre\-images of
$\ket{\alpha_j}$"=kets \cite{br1}). They should be read as the
detector responses at a collider, the interferometer flashes,
\thelike. It is these $\ket{\alpha_j}\ne\ket{\alpha_k}$ that
implement the empirical distinguishability of $\state\alpha$"=clicks
$\state\alpha_j\not\approx\state\alpha_k$ from each other, and
accumulation thereof into numerical arrays is formalized by the
objects $\frak a_j\in\bbC$.

The statistical weights $\{\fr_j\}$ is an observable entity and,
hence, the $\scr A$"=bases have a conventional terminology""---the
basis of an observable $\scr A$ with the eigen states
$\ket{\alpha_j}$. The presence of a numerical concept $\{\fr_j\}$ is
an integral part of the superposition theorem \cite{br1}. Otherwise,
we would have `an abstract bare' \lvs, and all the other and the
familiar theoretical structures would have had to be postulated in
their own rights. This would run counter to the task of ascertaining
the nature of quantum axioms. Interrelation between $\{\fr_j\}$ and
bases of quantum \lvs\ may be set forth more mathematically.

Being a collection of experimental numbers, the statistics
$\{\fr_j\}$ may have a source in \lvs\ (`to be calculated from')
only from its primary numeric quantities. But `bare' \lvs\ contains
nothing but vectors, field $\bbC$, and dimension $\dim\bbH$.
Therefore, for quantum observations these quantities may only be the
$\bbC$"=coefficients of superpositions
\begin{equation}\label{PsiE}
\frak c_1\bcdot\ket{\boe_1}\+
\frak c_2\bcdot\ket{\boe_2}\+\cdots=\ketPsi\;.
\end{equation}
The meaningfulness/"!uniqueness of the $\frak c$'s, for a given vector
$\ketPsi$, is possible only if the family
$\{\ket{\boe_1},\ket{\boe_2},\ldots\}$ forms a basis of linearly
independent vectors. Coefficients $\frak c$'s in the arbitrary
abstract superpositions
\begin{equation}\label{super}
\frak c_1\bcdot\ket{\Phi_1}\+ \frak c_2\bcdot\ket{\Phi_2}
\end{equation}
would not do for this.

Since there are no restrictions on the bases of \lvs, the
$\frak c$"=coefficients in \eqref{PsiE}--\eqref{super} may be any
collections; \ie, any coordinates may be assigned to every vector.
Hence, formula for computation of frequencies $\{\fr_j\}$ from
coefficients $\frak c_j$ may exist only if $\{\ket{\boe_j}\}$ is a
basis but it is not free. What is required is a basis, for which
both its ket"=vectors and the expansion coefficients in \eqref{PsiE}
keep the nature of the origin of the basis as a
concept""---""accumulating the (in)distinguishable micro"=events
$\state\alpha_j$. Therefore it is permissible to rely only on
coefficients $\frak a_j$ in $\ket\alpha$"=expansions \eqref{=}, \ie,
on bases, to which the $\fr$"=statistics has been attached. This is
implemented by a core object---the statistical length \cite{br2}
\begin{equation}\label{SL}
\StatL\big(\frak a_1\bcdot\ket{\alpha_1}
\+ \frak a_2\bcdot\ket{\alpha_2}\+\cdots\big)\;.
\end{equation}

The function
$\StatL(\cdot{\cdot}\cdot)\FED\cal N[\cdot{\cdot}\cdot]$ is utterly
minimalistic in its derivation""---it requires neither Hilbert'
space nor physics, is unique, and has a sum"=of"=squares form
\cite{br2}
\begin{equation}\label{a2}
\cal N\big[\frak a_1\bcdot\ket{\alpha_1}
\+ \frak a_2\bcdot\ket{\alpha_2}\+\cdots\big]=|\frak a_1|^2+
|\frak a_2|^2+\cdots\;.
\end{equation}
From this, there immediately follows the formula
\begin{equation}\label{BR}
\fr_k=\frac{|\frak a_k|^2}{|\frak a_1|^2+|\frak a_2|^2+\cdots}\;,
\end{equation}
which will be further turned into the standard writing
$\fr=\mbig[1]|\langle\bo\psi|\bo\phi\rangle\mbig[1]|^2$ of the
famous Born rule~\cite{born}. Expressed another way, the formula to
be directly deduced is not a Neumann--Hilbertian version of this
rule like $\fr=|\skew{-1}\widehat{\msf P}\bo\psi|^2$, but rather the
purely numerical full-ratio~\eqref{BR}. The derivation exploits
\cite{br2} only the minimal structure---the vector space, but
numerical form \eqref{BR} should be represented invariantly.

In consequence, whereas the concept of a quantum $\scr A$"=basis has
been present in the superposition theorem \cite{br1} and has been
exploited when deriving \eqref{BR}, formalization of this term has
yet to be created.

\subsection{$\|\bo\psi\|$ and
$\langle\bo\psi|\bo\varphi\rangle$}\label{22}

As for the scalar product and norm on \lvs, the self"=evident usage
of these structures seems also illogical in keeping with the
axiom-free building the \qt. At the moment, they have no empirical
grounds. For example, empiricism of experiment is inherently unable
to tell us anything about property
$\langle\bo\psi|\bo\varphi\rangle=\langle\bo\varphi|\bo\psi\rangle^*$
for a certain map $(\bbH\Times\bbH)
\mathrel{\Over{\sss\langle\bcdot|\bcdot\rangle}{\longmapsto}} \bbC$;
this point is often discussed in quantum logic \cite{engesser}.

The habitual classical and physically illustrative argumentation is
not an exception. More to the point, as noted above, such
argumentation has actually been banned \cite{br1}, because it is a
source of confusions and of logical inconsistencies
\cite{silverman}. Say, the typical use of the aforementioned map
$|\langle\texttt{in}|\texttt{out}\rangle|^2$ for calculation of the
`transition probabilities between states $\ket{{}\texttt{before}}$ and
$\ket{{}\texttt{after}}$' should be regarded in a very conditional way,
since each of the words (and combinations thereof) in this sentence
is still a subject of discussion \cite{laloe, hren, aaronson+,
fuchs2}, especially the philosophical.

If, in contrast, we draw on these structures as on the ready"=made
ones, then one will be required the guessing and substantiating
their quite specific properties such as parallelogram rule and the
triangle inequality
\begin{equation*}
\|\bo\psi \+ \bo\varphi\|^2 +
\|\bo\psi\mathbin{\hat{\smash-}}\bo\varphi\|^2= 2\,\|\bo\psi\|^2 +
2\,\|\bo\varphi\|^2\;,\qquad \|\bo\psi\| + \|\bo\varphi\|\geqslant\|\bo\psi \+
\bo\varphi\|\;,
\end{equation*}
the polarization procedure
\begin{equation*}
4\,\langle\bo\psi|\bo\varphi\rangle=\|\bo\psi \+ \bo\varphi\|^2 -
\|\bo\psi\mathbin{\hat{\smash-}}\bo\varphi\|^2 + \ri\,\|\bo\psi \+
\ri\,{\bcdot}\,\bo\varphi\|^2 -
\ri\,\|\bo\psi\mathbin{\hat{\smash-}}\ri\,{\bcdot}\,\bo\varphi\|^2\;,
\end{equation*}
the companion von~Neumann--Jordan--Fr\'echet theorems, equivalence
of norms on \lvs, \thelike\ \cite{fried, halm, gudder}. The
questions ``whence/"!why?\@'' do not go away here (just being shifted to
another domain), and the familiar quantum `difficulties in
relationship' \ceils{physics $\rightleftarrows$ mathematics} are
aggravated inasmuch as the motivating is substituted for definienda
and the inferencing""---for proofs. Beyond the matter of quantum
bases, the typical example is a slightly disconcerting Riesz'
theorem on the representation of a (bounded) linear functional by a
scalar product \cite{fried, gudder}. Functionals on the \lvs\ are
more primary, as they do not require for their own definition
anything but \lvs\ itself, whilst the scalar product is a quite
nontrivial external add-on over it. Not to wonder why and whether
the concept of a linear functional should come into play (see
sect.~\ref{HS}).

\subsection{Summary}\label{resume}

Now, the logic in foundations of \qt\ should, ideally, avoid the
sequencing \eqref{phil-}, and we will adhere to the scheme back
toward \eqref{phil-}:
\begin{gather*}
\text{\big(\ceils{common `philosophical background'}\;\tplus\;\big)}\\
\text{\ceils{motivation} $\rightarrowtail$ \ceils{inference}
$\rightarrowtail$ \ceils{construct} $\rightarrowtail$
\ceils{formalization, terminology}}\;.
\end{gather*}
This is because the search precisely for such a scheme
\cite[Introduction~(!)]{fuchs2} constitutes the dominant bulk of the
subject matter of so extensive literature on math of quantum
foundations \cite{engesser, silverman, dAriano+}. That is why nearly
all the quantum terminology we have been accustomed \cite{fried,
gudder}, \eg,
\begin{gather*}
\text{\ceils{scalar product}, \quad\ceils{norm}, \quad\ceils{dual
space}}, \quad\\\text{\ceils{linear operators/functionals},\quad
\ceils{self-adjointness},\quad\ceils{spectra}}\;,
\intertext{expressions and devices like}
\text{\ceils{$\langle\bo\psi|\bo\varphi\rangle$,\quad
$\skew{-1}\widehat{\msf P}$,\quad
$\frak a_1\frak b_{\smash{1}}^* +
\frak a_2\frak b_{\smash{2}}^* + \cdots$,\quad
\ldots}}
\end{gather*}
should be viewed methodologically as \emph{non}"=existing at the
beginning. The usage of such terms is forbidden until they have been
created explicitly without their implication a~priory and without
`guessing or fitting underneath' the familiar constructs.

A few words about the \lvs\ itself. As was the case with deriving
the rule \eqref{BR}, the whole discourse does not depend on
``mechanism for the emergence of \lvs'' \cite[p.~4]{br2}. If desired,
the statement about quantum linearity \cite{br1} may be
forgotten/"!ignored as a theorem, and one may keep within the
orthodox/"!postulational view on the superposition. Then the questions
are:
\begin{itemize}
\item Is such a non-Hilbertian \lvs-version of the superposition
 principle sufficient to produce the remaining (counterintuitive for
 comprehension) quantum math? Will that minimalism call for extra
 premises or axioms?
\end{itemize}
The answers are `yes and no', respectively, as soon as this \lvs\ is
supplemented with the `observable' function~\eqref{a2}.

Given what we have to work with, the problem as to how a calculus on
the $\bbH$"=space will look like, \ie\ which mathematical structures
are entailed by quantum paradigm, is determined in our exposition
\emph{only} by the data pair
\begin{equation}\label{scheme}
\text{\ceils{vector space $\bbH$}\quad\tplus\quad
\ceils{the number function \eqref{a2}}}
\end{equation}
(and nothing more). In other words, mathematics of \qt\ should be
created not from axioms, calculus acquires the character of a
linear"=algebra calculus \cite{halm, fried}, but emergence of the
structures""---""operators, their spectra, hermicity,
quadratic/"!bilinear forms, the informational conceptions, quantum
dynamics, \thelike---is not postulated a~priori. None of these
terminologies will be further required.

The numerical nature of Born's rule---""function \StatL\ on abstract
\lvs---has already been described in \cite{br2}, and we may
therefore characterize subsequent actions also in the `vein of the
numerical ideology'. This is because the abstract linear space of
the $\ketPsi$"=vectors is isomorphic \cite{halm, fried} to its model
of the number collections $(\frak a_1,\frak a_2,\ldots) \in \bbC^{\sss
N}$:
\begin{enumerate}
\item[\hypertarget{1}{$\red1^{\Circ}$}] The numerical form of relations
 in $\scr A$"=bases. Orthogonality (sect.~\ref{sec-ort}).

\item[\hypertarget{2}{$\red2^{\Circ}$}] The numerical form of a new
 object""---""product of vectors (sect.~\ref{sc}).

\item[\hypertarget{3}{$\red3^{\Circ}$}] Axiomatization of the scalar
 product $\langle\bo\Psi|\bo\Phi\rangle$ (sect.~\ref{Formal}).

\item[\hypertarget{4}{$\red4^{\Circ}$}] Revision of orthogonality and of
 length in classical geometry (sect.~\ref{Pif}).

\item[\hypertarget{5}{$\red5^{\Circ}$}] Back to the abstraction~$\bbH$.
 The unitary \lvs.

\item[\hypertarget{6}{$\red6^{\Circ}$}] Topology and norm. The quantum
 Hilbert space (sect.~\ref{topology}).
\end{enumerate}
Thus, the appearance sequence of what follows, \ie
\begin{gather*}
\text{\ceils{statistics} $\rightarrowtail$ \ceils{\lvs"=bases}
$\rightarrowtail$ \ceils{unitarity} $\rightarrowtail$
\ceils{`orthogonality'} $\rightarrowtail$ \ceils{observables}
$\rightarrowtail$}\\
\text{\ceils{scalar product}
$\rightarrowtail$\ceils{axiomatization} $\rightarrowtail$
\ceils{unitary space} $\rightarrowtail$ \ceils{topology}
$\rightarrowtail$ \ceils{Hilbert space}}
\end{gather*}
is pretty much the opposite of the typical \qt"=axiomatics followed
by interpreting: \ceils{axioms/"!math of Hilbert space}
$\rightarrowtail$ \ceils{(statistical) treatment}.

\section{What is (quantum) orthogonality?}\label{sec-ort}

Our next immediate task is to find out a formulation of
distinguishability of $\scr A$"=bases from the others, and in
addition to the \lvs\ itself, the \emph{only thing} we may exploit
is the additive nature of the statistical"=length function
\eqref{a2}:
\begin{equation}\label{Naj}
\cal N\big[\frak a_1\bcdot\ket{\alpha_1}
\+ \frak a_2\bcdot\ket{\alpha_2}\+\cdots\big]
=\cal N[\frak a_1\bcdot\ket{\alpha_1}]+
\cal N[\frak a_2\bcdot\ket{\alpha_2}]+\cdots\qquad
\forall\,\frak a_j\;.
\end{equation}
It is admitted only to such bases or, to be precise, it is through
this nature that determines them \cite{br2}. One notices that we do
not even assume a~priori that the belonging to a basis of an
observable should be numerical in character and binary, \ie, that it
should be given by some numerical relation between its two members.

At the same time, additivity \eqref{Naj} and arbitrariness of
coefficients $\frak a_j$ say that the $\cal N$"=function is defined,
as a minimum, by a sum of only \emph{two} terms. The general case
\eqref{Naj} is processed recursively. Therefore the question of
belonging to $\scr A$"=basis reduces to the pairwise relations
between its elements. This gives rise to the term binary
superstructure in theory.

In order not to burden notation with indices, let us write
\begin{equation}\label{=2}
\cal N\big[a\bcdot\ket\alpha \+ b\bcdot\ket\beta\big]
=\cal N[a\bcdot\ket\alpha]+ \cal N[b\bcdot\ket\beta]\qquad \forall\,
a, b\in\bbC\;,
\end{equation}
assuming that vectors $\ket\alpha$ and $\ket{{}\bo\beta}$ belong to
a certain $\scr A$"=basis.

\subsection{The nature of orthogonality}

In line with ideology
\hyperlink{1}{$\red1^{\Circ}$}--\hyperlink{6}{$\red6^{\Circ}$}, let
us switch over a $\bbC^{\sss N}$"=number equivalent of the
space~$\bbH$. That is, in accord with \eqref{=}, we identify all its
vectors with their number $\scr A$"=representatives:
\begin{equation}\label{HHA}
\lceil\text{vectors }
\ketPsi\in\bbH\rceil\quad\mathrel{\Over{\scr A}{\longmapsto}}\quad
\lceil\text{collections }(\frak a_1,\frak a_2,\ldots)\in\bbC^{\sss
N}\rceil\FED\bbH_{\sss\scr A}\;.
\end{equation}
Because of isomorphism between $\bbH$ and $\bbC^{\sss N}$
\cite{fried}, the algebraic operations on the space
$\bbH_{\sss\scr A}$ will be denoted by the same symbols $\{\+,
\bcdot\}$ :
\begin{equation}\label{algebra}
(\frak a_1,\frak a_2,\ldots)\+(\frak b_1,\frak b_2,\ldots)\DEF
(\frak a_1+\frak b_1,\frak a_2+\frak b_2,\ldots)\,,\qquad c\bcdot
(\frak a_1,\frak a_2,\ldots)\DEF(c\,\frak a_1,c\,\frak a_2,\ldots)\;.
\end{equation}

The statistical"=length function $\cal N$, as a function of an
$\scr A$"=representation of vector \eqref{SL}, turns into a function
$\cal N_{\!\!\!\sss\scr A}$ that is fully defined on all the
$\bbH_{\sss\scr A}$"=vectors by formula
\begin{equation*}
\cal N_{\!\!\!\sss\scr A}\big[(\frak a_1,\frak a_2,\ldots)\big]=
|\frak a_1|^2+|\frak a_2|^2+\cdots\qquad\forall\,
(\frak a_1,\frak a_2,\ldots)\in \bbH_{\sss\scr A}\;;
\end{equation*}
now, without reference to the concept of a basis. But since the
$\cal N$ was being derived as an $\scr A$"=invariant
construct""---this is one of its axioms \cite{br2}---the change of
bases\footnote{The `change of bases and representations', as a
mathematical action, is involved not just for a formalization, but
is a fundamental point for the emergence of \qt\ itself
\cite[sect.~5.4]{br1}.} should be carried over to the space
$\bbH_{\sss\scr A}$. This amounts to a transition to the same but
one more space~$\bbH_{\sss\scr B}$. To put it differently,
invariance speaks about a consistent universality of the squares
formula
\begin{equation}\label{Nb}
\cal N_{\!\!\!\sss\scr B}\big[(\frak b_1,\frak b_2,\ldots)\big]=
|\frak b_1|^2+|\frak b_2|^2+\cdots\qquad\forall\,
(\frak b_1,\frak b_2,\ldots)\in \bbH_{\sss\scr B}\;,
\end{equation}
\ie, about the equality
$\cal N_{\!\!\!\sss\scr A}\big[(\frak a_1,\frak a_2,\ldots)\big]=
\cal N_{\!\!\!\sss\scr B}\big[(\frak b_1,\frak b_2,\ldots)\big]$, when
$(\frak a_1,\frak a_2,\ldots)\in\bbH_{\sss\scr A}$ and
$(\frak b_1,\frak b_2,\ldots)\in\bbH_{\sss\scr B}$ do represent the one
vector $\ketPsi\in\bbH$. Then one may forget the statistical
treatment to the $\scr A$"=expansions \eqref{=} and regard the
$\cal N$ as an abstract and well"=defined function on all the
spaces~$\bbH_{\sss\scr A}$.

The two $\scr A$"=base vectors $\ket\alpha$ and $\ket\beta$ in
\eqref{=2} play a dedicated role at the moment. Their coordinates
have special form
\begin{equation}\label{101}
\begin{array}{lcl@{}r}
\bbH\ni\ket\alpha&\mathrel{\Over{\scr A}{\longmapsto}}&
(1,0,0,\ldots)&{}\in\bbH_{\sss\scr A}\;,\\[1ex]
\bbH\ni\ket\beta&\mathrel{\Over{\scr A}{\longmapsto}}&
(0,1,0,\ldots)&{}\in \bbH_{\sss\scr A}\;.
\end{array}
\end{equation}
What can be said about their representations in other
$\bbH_{\sss\scr A}$'s? Let their $\scr B$"=representatives, after
transition to a different $\scr A$"=basis $\scr B$, be designated as
\begin{equation*}
\begin{array}{lcl@{}r}
\bbH\ni\ket\alpha&\mathrel{\Over{\scr B}{\longmapsto}}&
(\frak A_1,\frak A_2,\ldots)&{}\in\bbH_{\sss\scr B}\;,\\[1ex]
\bbH\ni\ket\beta&\mathrel{\Over{\scr B}{\longmapsto}}&
(\frak B_1,\frak B_2,\ldots)&{}\in\bbH_{\sss\scr B}\;.
\end{array}
\end{equation*}
These have already a considerable numerical freedom $\frak A_1,\ldots,
\frak B_1,\ldots\in\bbC$.

Invoke now the additivity, \ie, let us write down \eqref{=2} in this
new basis:
\begin{equation*}
\cal N_{\!\!\!\sss\scr B}\big[a\bcdot(\frak A_1,\frak A_2,\ldots)
\+b\bcdot(\frak B_1,\frak B_2,\ldots)\big]=
\cal N_{\!\!\!\sss\scr B}\big[a\bcdot(\frak A_1,\frak A_2,\ldots)\big]+
\cal N_{\!\!\!\sss\scr B}\big[b\bcdot(\frak B_1,\frak B_2,\ldots)\big]\;.
\end{equation*}
According to algebra \eqref{algebra}, we have
\begin{equation*}
\cal N_{\!\!\!\sss\scr B}\big[(a\,\frak A_1+b\,\frak B_1,
a\,\frak A_2+b\,\frak B_2,\ldots)\big]=
\cal N_{\!\!\!\sss\scr B}\big[(a\,\frak A_1,a\,\frak A_2,\ldots)\big]+
\cal N_{\!\!\!\sss\scr B}\big[(b\,\frak B_1,b\,\frak B_2,\ldots)\big]\;,
\end{equation*}
and according to the `square rule' \eqref{Nb}, we obtain the
equality
\begin{multline*}
(a\,\frak A_1+b\,\frak B_1)(a\,\frak A_1+b\,\frak B_1)^*+
(a\,\frak A_2+b\,\frak B_2)(a\,\frak A_2+b\,\frak B_2)^*+\cdots\\
=\big\{(a\,\frak A_1)(a\,\frak A_1)^*+(a\,\frak A_2)(a\,\frak A_2)^*+
\cdots\}+
\big\{(b\,\frak B_1)(b\,\frak B_1)^*+(b\,\frak B_2)(b\,\frak B_2)^*+
\cdots\}\;.
\end{multline*}
By expanding and canceling, one arrives at equation (recalling the
arbitrariness of $a,b$)
\begin{equation*}
(a\,b^*)\,(\frak A_1^{}\,\frak B_1^*+\frak A_2^{}\,\frak B_2^*+\cdots)+
(a^*\,b)\,(\frak A_1^*\,\frak B_1^{}+\frak A_2^*\,\frak B_2^{}
+\cdots)=0\qquad \forall\, a,b\in\bbC\;,
\end{equation*}
that is
\begin{equation*}
\msf{Re}\big\{c\,(\frak A_1^{}\,\frak B_1^*+
\frak A_2^{}\,\frak B_2^*+\cdots)\big\}=0\qquad \forall\, c\in\bbC\;.
\end{equation*}
It follows that
\begin{itemize}
\item for every two $\scr A$"=base vectors $\ket\alpha$ and $\ket\beta$,
 their coordinates in a basis of any other observable must satisfy
 the numeric relation
\begin{equation}\label{ort}
\frak A_1^{}\,\frak B_1^*+\frak A_2^{}\,\frak B_2^*+\cdots=0\;.
\end{equation}
\end{itemize}
Whilst this relation is coordinate, its meaning is absolute for all
the $\scr A$"=bases, and we call it \emph{orthogonality relation}
$\ket\alpha\mathrel\bot\ket\beta$. The
point~\hyperlink{1}{$\red1^{\Circ}$} has been completed. The nature
of orthogonality as of a concept is identical to that of Born's
rule---the numerical.

\subsection{\ceils{observable} $=$
\ceils{orthogonal basis}}\label{basis}

In view of the fact that the associating a \StatL\ with expansions
\eqref{=} is the only thing that distinguishes the quantum space
from merely linear $V$, the mechanism for making an abstract \lvs\
the quantum $\bbH$ \emph{is not axiomatic} and is as follows.

One considers an abstract $V$. Any vector $\ket{\boe_1}\in V$ may be
viewed as an eigen one for certain observable $\scr A_{\Circ}$; this
is fully aligned with the statistical genesis of ket"=objects
\cite[sect.~6.2]{br1}. Let us declare this vector to be an
$\ket\alpha$"=vector for the~$\scr A_{\Circ}$. Similarly for any
other linearly independent vectors $\{\ket{\boe_2},\ldots\}=
\{\ket{\alpha_2},\ldots\}$ down to exhausting the dimension $\dim V$.
One obtains an \lvs\ with a \emph{dedicated} (`good') basis---the
(orthogonal) basis of a certain (any) observable~$\scr A_{\Circ}$.
Such an act does always have a set"=theoretic character and always is
a matter of the declaration/"!appointment. Despite the title
`observable' in this appointment, there is no sense in looking for
its origin in the temporal processes like \ceils{quantum
$\rightarrowtail$ classical} or for a physical underlying reason
through the natural"=language meaning to the word `observable'.
Similarly, the geometric `meaning' of the word orthogonal. An
analogy: in a set, no its element is `innately good/"!bad' (say, by
physical reasons) until over this set there has been created a math
superstructure"=criterion, according to which some set"=members differ
from the others. The role of a superstructure over \lvs\ is played
here by the very basis $\scr A_{\Circ}$ with the ascribed function
\StatL\ \eqref{SL}--\eqref{a2}, and in which the orthogonality
property \eqref{ort} is seen to be `hard"=wired'. Orthogonality for
$\scr A_{\Circ}$"=elements does automatically hold due to
\eqref{101}.

Let us carry out all possible transformations $\cal U$ of
$\scr A_{\Circ}$ that preserve the property `to be an
$\scr A$"=basis'. Emergence of the
$\cal U$"=transformations\footnote{Unitarity $\cal U$ at the moment
is (and arises originally as) but a property of a numerical
collection $\cal U_\textit{jk}$, \ie, it is not an algebraic object and
is not of the invariant nature.} has been described in \cite{br2}.
One gets all the other $\scr A$"=bases and, thereby, they become
special (not arbitrary) since the $\cal U$"=matrices are not
arbitrary:
\begin{equation}\label{U*}
\cal U^\T \cal U^*=\mathds1\;.
\end{equation}
These may be normally referred to as complex"=orthogonal. Due to
group property, the transformations $\cal U$ `depersonalize' the
initial $\scr A_{\Circ}$, and it ceases to be a dedicated basis or,
according to the physical terminology, the preferable one
\cite[``pointer state'']{busch}. Such $\scr A_{\Circ}$"=bases---an
oft"=discussed subject in quantum measurement theory
\cite{busch}---should not be present in \qt\ \cite[sect.~6a]{br2},
because this has been demanded by the basic principle of invariance
and a prohibition on the physical meaning to
$\ket{{}\texttt{ket}}$"=vectors \cite{fuchs}. The remaining bases
$\{\ket{{}\textbf{e}_j}\}$ drop out of the $\cal U$"=series---they are
not orthogonal, non-$\scr A$"=bases---and expansions \eqref{PsiE}
over them are `bad' in the context of the statistical function
$\cal N$; its values are ill"=defined. The $\frak c_j$"=numbers in
\eqref{PsiE} simply do not have any observable meaning, just like it
is not in arbitrary $(\+)$"=superpositions \eqref{super}.

As an outcome, the concept of quantum observable""---we are not
talking about its numerical values $\alpha_j$ ---is identical at the
moment to the orthogonal basis, and vice versa; though the term
``orthogonal'' has not yet been completed as a mathematical structure.
By nature, such a meaning to the $\scr A$ does not require the
structure of a self"=adjoint operator, its real spectrum
$\{\alpha_j\}$, the eigen"=value problem, and Hilbertian space.
Getting off the subject slightly, all the quantum commutatives
$\{\hA, \hB, \ldots\}$ may be thought in effect of as the same
observable $\scr A$ but with different numerical values for these
spectral $\alpha_j$"=labels that being assigned to members of a
common orthogonal base-set $\{\ket{\alpha_j}\}$. Here, no use is
required of a concept the operator. The orthogonality is discussed
also in sect.~\ref{Pif}.

\section{Whence the scalar product?}\label{sc}

If quantum $\ket\alpha$"=representations produce statistics, then
what do we multiply the abstract $\ketPsi$"=states for
\cite{engesser}? Why scalarly?

\subsection{Statement of problem. Why binarity?}

Quantum theory is a numerical one, which is why upon establishing
the properties of $\scr A$"=bases, it is necessary to address the
arbitrary vectors in order to build the formal calculus on~$\bbH$.
We still do not have it per~se and what it should consist of is
undefined at the moment.

Because the spectral labels of $\ket\alpha$"=vectors (spectra) and
linear operators are absent in discourse at the moment (they are not
required and have not yet come into being), we have deal merely with
$\frak a$"=numbers in expansions \eqref{=}. Being the carriers of
$\fr$"=statistics \eqref{BR}, these numbers will represent the
subject matter of the $\bbH$"=calculus, and it must be
independent""---put this in a definition of the word
$\bbH$"=calculus""---of the coordinate
spaces"=representations~$\bbH_{\sss\scr A}$.

The aforesaid means that the space of quantum states $\bbH$ is
supplemented with the task of deducing an invariant formula
\begin{equation}\label{F}
\frak a_j=\Over[2.2]{\smash{(?)}}{\bo\aleph}\!
\big(\ketPsi,\ket{\alpha_j}\big)
\end{equation}
as an operation of `extracting' the $\frak a$"=coefficient from
expansion \eqref{=} when $\ket{\alpha_j}$ is a member of an
$\scr A$"=basis. Invariance of the formula is also called for because
the $\bbH$"=space itself owes its existence to the availability of at
least two instruments $\scr A$ and $\scr B$, and none of them are
preferable. The formula is necessary also for obtaining the
invariant writing of Born's rule \eqref{BR}. Notice that any
different way of searching-for or the `right proof' of the rule will
knowingly have not met with success because, in experiment, there
are neither operators nor the
$\langle\texttt{bra}|\texttt{ket}\rangle$"=abstracta; there is no even the
arithmetic there \cite[sect.~2.3]{br1}.

Inasmuch as not only is the $\ketPsi$ arbitrary, but every vector of
the space may serve as an eigen one $\ket{\alpha_j}$, then the
problem \eqref{F} should be solved through a certain (yet another)
\emph{binary} superstructure over the entire~$\bbH$:
$(\bbH\Times\bbH)\mathrel{\Over[1.1]{\bo\aleph}{\mapsto}} \bbC$.
That is just what the theory of function $\bo\aleph$ is. It is
appended to the $\bbH$"=space, which remains as is without the need
for its (over/re)defining.

How to search for $\bo\aleph$? Has it had, being a function of
unnecessarily nonorthogonal vectors, a link with orthogonality of
the $\scr A$"=basis ones? To be totally precise, we do not even have
the concept of nonorthogonality, since orthogonality \eqref{ort} is
currently a structure not over the entire $\bbH$ but is an exclusive
property of some special collections""---bases of observables.

\subsection{Product of arbitrary vectors}

While remaining within `numerical ideology'
\hyperlink{1}{$\red1^{\Circ}$}--\hyperlink{6}{$\red6^{\Circ}$}, let
us write an equality of the two $\scr A$"=representations for
$\ketPsi$:
\begin{equation}\label{AB}
\frak a_1\bcdot\ket{\alpha_1}\+
\frak a_2\bcdot \ket{\alpha_2}\+\cdots=\ketPsi=
\frak b_1\bcdot\ket{\beta_1}\+
\frak b_2\bcdot \ket{\beta_2}\+\cdots\;.
\end{equation}
To solve the task \eqref{F}, one suffices to solve it at first in
coordinates, \ie, to express coordinates $\frak a_j$ through
coordinates $\{\frak b_j\}$ of arbitrary vector $\ketPsi$ in any new
$\scr A$"=basis $\{\ket{\beta_j}\}_{\!\!\sss\scr B\mathstrut}$. Having obtained
an answer, this is the pt.~\hyperlink{2}{$\red2^{\Circ}$}, we come
back from the coordinate language of spaces \{$\bbH_{\sss\scr A}$,
$\bbH_{\sss\scr B}$, \ldots\} to the initial $\bbH$; this will be done
in sect.~\ref{Formal} as required by
pts.~\hyperlink{3}{$\red3^{\Circ}$}
and~\hyperlink{5}{$\red5^{\Circ}$}.

By virtue of unitarity \eqref{U*}, the relationship between bases
$\{\ket{\alpha_j}\}$ and $\{\ket{\beta_k}\}$ is as follows:
\begin{equation}\label{aUb}
\begin{aligned}
\ket{\beta_1}&=\cal U_{11}\bcdot\ket{\alpha_1}\+
\cal U_{12}\bcdot\ket{\alpha_2}\+\cdots\qquad\quad&
\ket{\alpha_1}&=\cal U_{11}^*\bcdot\ket{\beta_1}\+
\cal U_{21}^*\bcdot\ket{\beta_2}\+\cdots\\
\ket{\beta_2}&=\cal U_{21}\bcdot\ket{\alpha_1}\+
\cal U_{22}\bcdot\ket{\alpha_2}\+\cdots\,&
\ket{\alpha_2}&=\cal U_{12}^*\bcdot\ket{\beta_1}\+
\cal U_{22}^*\bcdot\ket{\beta_2}\+\cdots\\[0mm]
\text{\makebox[0em][l]{$\cdots\cdots\cdots\cdots\cdots\cdots$}}&&
\text{\makebox[0em][l]{$\cdots\cdots\cdots\cdots\cdots\cdots$}}
\end{aligned}\quad.
\end{equation}
Properties of unitary matrices include the algebraic relations
between $\cal U_\textit{jk}$ and $\ds \cal U_\textit{jk}^*$
\cite[sect.~IX.7]{gant}, but we will view of them also as
non"=existent. However, it is clear that they are corollaries of
formulas \eqref{101}, \eqref{U*}, and of the unit statistical length
\begin{equation}\label{U1}
\cal N[\ket{\alpha_j}]\quad\rightarrowtail\quad
\cal N_{\!\!\!\sss\scr A}[(0,\ldots,0,1,0,\ldots)]=
\cal N_{\!\!\!\sss\scr B}[\ldots]=\cdots=1\;.
\end{equation}

Substitution of $\ket{\beta_j}$ from \eqref{aUb} into \eqref{AB}
gives the commonly known rule of conversion between the $\frak a_j$-
and $\frak b_j$"=coordinates of one and the same vector $\ketPsi$:
\begin{equation}\label{temp}
\begin{aligned}
\frak a_1&=\frak b_1\,\cal U_{11}+\frak b_2\,\cal U_{21}+\cdots\\
\frak a_2&=\frak b_1\,\cal U_{12}+\frak b_2\,\cal U_{22}+\cdots\\[0mm]
\text{\makebox[0em][l]{$\cdots\cdots\cdots\cdots\cdots\cdots$}}
\end{aligned}\quad.
\end{equation}
Right hand part of these formulas needs to be realized in the
$\bbH_{\sss\scr B}$"=space. Here, according to the task \eqref{F},
there must appear only the vectors $\ketPsi$ and $\ket{\alpha_j}$;
more precisely, coordinates of these objects. Is this possible?

For simplicity, consider the 1"~st formula in \eqref{temp}. In it,
the row $(\frak b_1,\ldots)$ is already an $\bbH_{\sss\scr B}$"=image of
the first $\bo\aleph$"=argument in \eqref{F}, \ie\ $\ketPsi$:
\begin{equation*}
\eqref{AB}\quad\hence\quad\ketPsi\mathrel{\Over{\scr B}{\longmapsto}}
(\frak b_1,\frak b_2,\ldots)\FED\msf\Psi\in \bbH_{\sss\scr B}\;,
\end{equation*}
whereas the row $(\cal U_{11},\cal U_{21},\ldots)$, on the face of it,
is a free aggregate of side numbers. However, according to the
second module in representations \eqref{aUb}, one has
\begin{equation}\label{HaB}
\ket{\alpha_1}\mathrel{\Over{\scr B}{\longmapsto}}
(\cal U_{11}^*,\cal U_{21}^*,\ldots)\FED\msf A\in \bbH_{\sss\scr B}\;,
\end{equation}
\ie, the aggregate $(\cal U_{11},\cal U_{21},\ldots)$ is exactly the
complex conjugation of coordinates of the second
$\bo\aleph$"=argument in \eqref{F}---""coordinates of vector
$\ket{\alpha_1}$ in basis~$\scr B$. Thus the right hand side of
formulas \eqref{temp}
\begin{equation*}
\frak b_1^{}\,(\cal U_{11}^*)^*+\frak b_2^{}\,(\cal U_{21}^*)^*+\cdots=
\cdots
\end{equation*}
turns, as required, into the coordinate expression \emph{of vectors}
(\ie\ $\cal U$ disappears):
\begin{equation*}
\cdots=\msf\Psi_1^{}\,\msf A_1^*+\msf\Psi_2^{}\,\msf A_2^*+\cdots\qquad
(=\frak a_1\text{ in $\scr B$"=representation})\;.
\end{equation*}
In its turn, this very expression coincides with the orthogonality
structure \eqref{ort}. With that, by contrast to \eqref{ort}, there
is nothing special about vectors $\msf\Psi,
\msf A\in\bbH_{\sss\scr B}$. Both of them may be arbitrary, with
only one restriction on \StatL\ of vector~$\msf A$:
\begin{equation}\label{UA1}
\eqref{U1}\quad\hence\quad
\cal U_{11}^*\,\cal U_{11}^{}+\cal U_{21}^*\,\cal U_{21}^{}+\cdots
=1\quad\hence\quad \msf A_1^{}\,\msf A_1^*+\msf A_2^{}\,\msf A_2^*+\cdots=1\quad
(=\cal N[\ket{\alpha_1}])\;.
\end{equation}

As a result, the vector $\ket{\alpha_1}
\mathrel{\Over[1.1]{\sss\scr A}{\mapsto}}
(1,0,\ldots)\in\bbH_{\sss\scr A}$, being a fixed one, has turned into
an arbitrary $\bbH_{\sss\scr B}$"=vector \eqref{HaB} thanks to the
freedom in~$\cal U$. In other words, both the orthogonality form
\eqref{ort} and the coordinate's calculation of any vector are
implemented in all the $\bbH_{\sss\scr B}$"=spaces through the unique
numeric expression $\ds\frak a_1^{}\frak b_1^*+\frak a_2^{}
\frak b_2^*+\cdots$. It becomes universal, and $\msf\Psi$ and
$\msf A$ may be replaced by the two arbitrary vectors $\msf A$,
$\msf B$:
\begin{equation*}
(\frak a_1,\frak a_2,\ldots)\FED\msf A\in\bbH_{\sss\scr A}\,,\quad
(\frak b_1,\frak b_2,\ldots)\FED\msf B\in\bbH_{\sss\scr A}\;.
\end{equation*}

This freedom allows us to forget that one of the vectors was a
preimage of the $\ket{\alpha_j}$"=eigen one in the map \eqref{HHA}
and consisted of zeroes/"!unities \eqref{101}. Even the restriction
\eqref{UA1} becomes fictitious since \StatL\ of any vector $\msf A$,
once it has been declared to be an $\scr A$"=basic one, can always be
scaled to the unity by \eqref{U1}. No orthogonality \eqref{ort} with
other $\ket\alpha$"=vectors gets lost by this.

\subsection{R\'esum\'e}

All the $\bbH_{\sss\scr B}$"=spaces are certain $\scr A$"=realizations
of the space~$\bbH$. Therefore, by virtue of arbitrariness of
$\{\frak a_j,\frak b_j\}$, we introduce a designation for the
$\bbC$"=numerical and new form"=`multiplication':
\begin{equation}\label{scalar}
\frak a_1^{}\frak b_1^*+\frak a_2^{} \frak b_2^*+\cdots\FED
(\msf A,\msf B)\,,\quad
\forall\, \msf A\,,\msf B\in\bbH_{\sss\scr A}\;.
\end{equation}
It becomes an $\bbH_{\sss\scr A}$"=representative of the invariant
binary construct $\bo\aleph$, now on the entire $\bbH$, and does
prepare solution to the problem \eqref{F}. Moreover, orthogonality
\eqref{ort} completely fits in the structure of this multiplication
when the value of this form vanishes. Any vectors $\msf\Psi$ and
$\msf A$ may now be multiplied according to the rule \eqref{scalar},
yielding their indecomposability $(\msf\Psi,\msf A)=0$ through each
other or, in accord with \eqref{F}, the expansion coefficients with
respect to orthogonal bases:
\begin{equation}\label{PsiA}
\msf\Psi=\frak a\bcdot\msf A\+
\lceil\text{$\msf A^{\!\!\sss\bot}$"=component}\rceil\,,\qquad
\frak a=\frac{(\msf\Psi,\msf A)}{(\msf A,\msf A)}\qquad
\forall\,\msf\Psi,\msf A\in\bbH_{\sss\scr A}\;.
\end{equation}
Orthogonality of basis $\{\msf A^{\!\!\sss(j)}\}$ may then be
consistently thought of as
$\ds(\msf A^{\!\!\sss(j)},\msf A^{\!\!\sss(k)})\sim \delta_\mathit{jk}$,
although it is, as before, independent of the construction
\eqref{scalar}.

The main conclusion now is a conclusion about the appearance
sequence. It is a weaker and a \emph{particular} structure of
orthogonality \eqref{ort}, not multiplication of \emph{arbitrary}
vectors \eqref{scalar}, that is primary when creating the quantum
mathematics.

In turn, the very orthogonality is preceded by the \StatL\
\eqref{a2}. Had we not stated the task \eqref{F}, the property
\eqref{ort} and unitarity would still exist as a property `to be an
$\scr A$"=basis'. Orthogonality does not depend on the task \eqref{F}
and `knows nothing' about calculating the $\frak a$"=coefficients in
expansions \eqref{=} by formulas like \eqref{PsiA} or much less by
the familiar versions thereof through Neumann's projectors
$\skew{-1}\widehat{\msf P}_{\![4]\sss\ket\alpha}$. When following a
different sequence, the \emph{formal} postulation of quantum"=vector
multiplication will always call for motivation of binarity, of
scalarness, and of linearity (see below); not to mention the
complex"=conjugation operation in axioms of the scalar product.

Certainly, all these commentaries can be carried over to the pure
mathematics""---""introduction of the additional math structures on
vector spaces at all \cite{gudder}. For example, S.~Gudder
introduces the concept of orthogonal additivity
\cite[sect.~5.2]{gudder}, whilst additivity \eqref{Naj} is primary
in itself \cite{br2} and orthogonality \eqref{ort} is its
consequence.

Generality of the multiplication-form \eqref{scalar} allows us to
cast away not only the restriction on \StatL\ of basis vectors
$\cal N_{\!\!\!\sss\scr A}[\msf A^{\!\sss(j)}]=1$ but even to forget the
concept itself. For all the $\scr A$"=instruments, the \StatL\ is
replaced now by the structure $(\msf A,\msf B)$ with redefinition
$\cal N_{\!\!\!\sss\scr A}[\msf A]= (\msf A,\msf A)$. Then formula
\eqref{BR}, having regard to \eqref{PsiA}, is obviously modified:
\begin{equation}\label{BR+}
\begin{aligned}
\fr_k&=
\frac{\ds\cal N_{\!\!\!\sss\scr A}[\frak a_k\bcdot\msf A^{\!\sss(k)}]}
{\ds\cal N_{\!\!\!\sss\scr A}[\frak a_1\bcdot\msf A^{\!\sss(1)}
\+\cdots]}= \frac{\ds|\frak a_k|^2\,
(\msf A^{\!\sss(k)},\msf A^{\!\sss(k)})}
{\ds\cal N_{\!\!\!\sss\scr A}[\msf\Psi]}\\
&=
\bigg|\frac{\ds(\msf\Psi,\msf A^{\!\sss(k)})}
{\ds(\msf A^{\!\sss(k)},\msf A^{\!\sss(k)})}\bigg|^2
\frac{\ds(\msf A^{\!\sss(k)},\msf A^{\!\sss(k)})}
{\ds(\msf\Psi,\msf\Psi)}=
\frac{\ds(\msf\Psi,\msf A^{\!\sss(k)})\,
(\msf A^{\!\sss(k)},\msf\Psi)}
{\ds(\msf A^{\!\sss(k)},\msf A^{\!\sss(k)})\,
(\msf\Psi,\msf\Psi)}\;.
\end{aligned}
\end{equation}
However we will remain within the framework $\cal N[\ket\alpha]=1$
in order not to replace the well"=established definition of
unitarity. See ref.~\cite{br4} for the full discussion on this unity
and the values $\cal N[|\bo\alpha\rangle]$'s.

\section{Space $\bbH^{\sss\langle\rangle}$}\label{Formal}

Let us take up the `returning' from $\{\bbH_{\sss\scr A},
\bbH_{\sss\scr B},\ldots\}$ to the $\ketPsi$"=abstracta of the space
$\bbH$; \ie\ pt.~\hyperlink{3}{$\red3^{\Circ}$}.

\subsection{Axiomatization}\label{Axiom}

The players of the construct \eqref{scalar} are a \emph{non}"=ordered
pair $\{\msf A, \msf B\}$ and the point that both of these vectors
are the \lvs"=algebra's elements. This is all we need when
formalizing the multiplication \eqref{scalar}.

The orthogonality relation is symmetric""---from
$\msf A\mathrel\bot\msf B$ there follows $\msf B\mathrel\bot\msf A$.
At the same time, the `equal quantum rights' of $\msf A$ and
$\msf B$ say that in expansion of one vector along a basis wherein
the second is its element, \ie\ in formulas
\begin{equation*}
\msf A=\frak c\bcdot\msf B\+
\lceil\text{$\msf B^{\sss\bot}$"=component}\rceil\,,\qquad
\msf B=\frak c'\bcdot\msf A\+
\lceil\text{$\msf A^{\!\!\sss\bot}$"=component}\rceil\;,
\end{equation*}
each of the vectors may be deemed first or second. We should
therefore consider a permutation of words \ceils{first
$\rightleftarrows$ second} in \eqref{scalar} and its
degeneration""---a case \ceils{the first} $=$ \ceils{the second}.
Obviously, the permutation $\frak a \rightleftarrows \frak b$
entails the properties
\begin{equation}\label{sc1}
(\msf A,\msf B)=(\msf B,\msf A)^*\;,\qquad\big\{(\msf A,\msf A)\geqslant0\,,\quad
(\msf A,\msf A)=0\;\hence\; \msf A=\msf 0\big\}\;.
\end{equation}
Notice that \eqref{sc1} is not merely a side property, but the one
that will define the axiomatization. It is the aforesaid
`equal-rights' that demand for this.

One is left with analyzing the algebraic operations $\{\+,\bcdot\}$
of the vector space itself:
\begin{equation*}
(\msf A\+\msf B,\msf C)={}?\,,\qquad
(\lambda\bcdot\msf A,\msf B)={}?\;.
\end{equation*}
Application of addition/"!multiplication \eqref{algebra} to
\eqref{scalar} gives the linearity rules:
\begin{equation}\label{sc2}
(\msf A\+\msf B,\msf C)=
(\msf A,\msf C)+(\msf B,\msf C)\;,\qquad
(\lambda\bcdot\msf A,\msf B)=
\lambda\times(\msf A,\msf B)\quad\forall\,\lambda\in\bbC\;.
\end{equation}
Algebra of \lvs\ does not contain any other structures such as the
order relation or continuity, which is why no quantum paradigm
requires any further steps for formalizing the
construct~\eqref{scalar}.

It is known that nonisomorphic models of the abstract $V$, under the
fixed dimension $\dim V$, do not exist \cite{fried, halm}. Therefore
we may forget about $\bbC^{\sss N}, \bbH_{\sss\scr A}$ and the word
`model' at all and turn the properties \eqref{sc1}--\eqref{sc2} for
$\{\bbH_{\sss\scr A}, \bbH_{\sss\scr B},\ldots\}$ into the abstract
axioms of a new abstract binary superstructure
$(\bbH\Times\bbH)\mapsto\bbC$.

\begin{comment}
Conversely, if a physical problem is realized by a certain model for
$\bbH$---""functional, the matrix"=based, by $\psi$"=functions,
\etc---\ie\ not necessarily by the $\bbC^{\sss N}$"=model, then the
model will in no way contradict the deduced axioms. At the same
time, the question about \emph{interpreting} the vectors of the
model must not appear, since identifying the wave functions
$\psi(\bo x,t)$ with some spatio"=temporal processes/"!phenomena will
be problematic \cite{silverman}, \cite[sect.~10.2]{br1}. Indeed, the
concepts of a field $\psi$ and of the very $(\bo x,t)$"=space are as
yet absent in theory. The real/"!experimental meaning is attached only
to the mod"=squares $|\frak a_k|^2$ in expansions \eqref{=}.

Non"=physicality of quantum states has been often discussed and
repeatedly pointed out in the literature \cite{ludwig5},
\cite[``quantum state does not represent an element of physical
reality'']{fuchs}. It is significant that the interpretation, in
mathematical logic \cite{frenkel}, is yet a further theory in its
own right. This is not a reconstruction or an `intrinsic reshuffle'
of the initial theory but the new mappings. In our setting, the
physical interpretation may arise only after the Born statistics
\eqref{BR}. These points are fully discussed in sect.~\ref{6th}.
\end{comment}

As a result, transforming the notation $(\msf\Psi,\msf\Phi) \goto
(\ketPsi,\ketPhi) \goto \langle\bo\Psi|\bo\Phi\rangle$, we claim
\begin{equation}\label{axiom}
\begin{array}{c}
\langle\bo\Psi\+\bo\Phi|\bo\Theta\rangle=
\langle\bo\Psi|\bo\Theta\rangle+
\langle\bo\Phi|\bo\Theta\rangle\;,\qquad
\langle\lambda\bcdot\bo\Psi|\bo\Phi\rangle=
\lambda\times\langle\bo\Psi|\bo\Phi\rangle\quad\forall\,\lambda\in\bbC\;,
\\[1ex]
\langle\bo\Psi|\bo\Phi\rangle=
\langle\bo\Phi|\bo\Psi\rangle^*\;,\qquad
{\big\{\langle\bo\Psi|\bo\Psi\rangle\geqslant0,\quad
\langle\bo\Psi|\bo\Psi\rangle=0\;\hence\;
\ketPsi=\ket0\big\}}
\end{array}
\end{equation}
and call construction $\langle\bo\Psi|\bo\Phi\rangle\in\bbC$ the
\emph{scalar/"!inner product} of vectors $\ketPsi$, $\ketPhi\in\bbH$.
The matrix unitarity \eqref{U*} may now be turned into an algebraic
object on \lvs, \ie, be redefined invariantly as a reversible
abstract transformation (operator) that preserves the scalar
product:
\begin{equation}\label{iso}
\bbH\mathrel{\Over{\widehat{\cal U}}{\longmapsto}}\bbH:\qquad
\big\langle\,\skew0\widehat{\cal U}\,\bo\Psi
|\,\skew0\widehat{\cal U}\,\bo\Phi\big\rangle=
\langle\bo\Psi|\bo\Phi\rangle\;.
\end{equation}

Upon supplementation of the space $\bbH$ with that product, \ie\
following the creation of the structure
\begin{equation}\label{H+}
\text{\ceils{$\bbH$"=space}\; \tplus\;
\ceils{\eqref{axiom}--\eqref{iso}}}\FED \bbH^{\sss\langle\rangle}\;,
\end{equation}
there arises a question about isomorphism of the
$\bbH^{\sss\langle\rangle}$"=space models. Whilst realization of the
abstract axioms \eqref{axiom} on one space $\bbH$ is already
multiple, the categoricity of structures \eqref{H+} under the fixed
dimension $\dim\bbH<\infty$ is well known \cite{halm, fried}. In
detail, for the two models $\bbH_1$, $\bbH_2$ of space $\bbH$, one
can move any basis of $\bbH_1$ into any other basis of $\bbH_2$ by a
one-to-one transformation. By virtue of this equivalence, every
orthogonal basis in $\bbH^{\sss\langle\rangle}_2$ has a preimage in
$\bbH^{\sss\langle\rangle}_1$, which is also an $\scr A$"=basis
there. Meanwhile, transition between the $\scr A$"=bases in $\bbH_1$
is a main property of unitarity $\cal U$---""invariance of \StatL.
Therefore, subsequent to bases, there transform uniquely both the
vectors and the values of all of the $\langle\bcdot |\bcdot
\rangle$"=products in $\bbH^{\sss\langle\rangle}_1$ and in
$\bbH^{\sss\langle\rangle}_2$. All the models for \eqref{H+} are
isomorphic.

\subsection{Hilbert space}\label{HS}

Now, solution to the problem \eqref{F} acquires the form
\begin{equation}\label{aj}
\frak a_j=\frac{\langle\bo\Psi|\bo\alpha_j\rangle}
{\langle\bo\alpha_j|\bo\alpha_j\rangle}\;,
\end{equation}
and the scalar product, as well as its isometry \eqref{iso}, may be
viewed as an invariant, yet auxiliary tool for calculating the
$\frak a$'s. It has no other motivation for emergence and it, along
with the known normalization requirement
$\langle\bo\Psi|\bo\Psi\rangle=1$, \emph{is not} a question about
inner axiomatics of the quantum $\ketPsi$"=state set. Incidentally,
it does \emph{not} appeal to the concept of a dual space and its
$\bra{{}\texttt{bra}}$"=vectors. These objects would require introducing
not only a certain linear function on $\bbH$ but also creating one
more (why, what for?\@) \lvs; a typical treatment of observables as
functionals on the first \lvs.

Terminology with new product can be continued by introducing the
projecting operation
\begin{equation*}
\frac{\langle\bo\Psi|\bo\alpha\rangle}
{\langle\bo\alpha|\bo\alpha\rangle}\FED\brak[,]{\bo\Psi}{\bo\alpha}\;,
\end{equation*}
and by rewriting the rule \eqref{aj}:
\begin{equation}\label{proj}
\frak a_j=\brak[,]{\bo\Psi}{\bo\alpha_j}\;.
\end{equation}
It is easy to see that the projection satisfies \eqref{axiom} except
for the transposition axiom:
\begin{equation}\label{linF}
\brak[,]{\bo\Psi\+\bo\Phi}{\bo\alpha}=
\brak[,]{\bo\Psi}{\bo\alpha}+
\brak[,]{\bo\Phi}{\bo\alpha}\,,\qquad
\brak[,]{\lambda\bcdot\bo\Psi}{\bo\alpha}=
\lambda\times\brak[,]{\bo\Psi}{\bo\alpha}\quad\forall\,\lambda\in\bbC\;.
\end{equation}
This is nothing but the axioms of the linear functional
$\msf P_{\!\!\!\!\smallket\alpha}[\ketPsi]=\brak[,]{\bo\Psi}{\bo\alpha}$
with a parameter $\ket\alpha$; this completes a comment on the Riesz
theorem mentioned in sect.~\ref{22}.

Let us reverse the discourse. For the invariant producing the
$\frak a$"=coefficient from \eqref{=}, one can easily get towards the
linearity properties \eqref{linF}. By this, however, the whole
theory of such a `functional $\msf P$"=calculus' would be
accomplished vacuously, for the formal linear functional
$\msf P_{\![7]\sss\ket{\boe}}[\ketPsi]$ `extracts'
$\frak c$"=coefficients from any expansions
\eqref{PsiE}--\eqref{super}. It contains no the idea of a quantum
basis, \ie, of a quantitative observability $\{\fr_j\}$, and
additivity of this $\msf P$ does not care even the linear dependence
of vectors. The idea, in contrast, is implemented through the
auxiliary and unary function \StatL, and \emph{nonlinear} at that.

Thus non"=axiomatic (without treatments and interpretations)
description of the quantum \lvs\ necessitates separating the notions
pertaining to the structural properties of the \qm"=state set $\bbH$
per~se from the \emph{calculus add-ons} over it. Attention should be
paid to the fact that the emergence of $\bbH^{\sss\langle\rangle}$,
usually perceived as a quantum analog of the classical phase"=space,
calls for neither `observable' Born's $\fr$"=numbers \eqref{BR} nor
even a physics that accompanies them. The space
$\bbH^{\sss\langle\rangle}$ \emph{is not a space of physical quantum
states} with familiar `illustrations' $\ket{{}\texttt{alive}} \+
\ket{{}\texttt{dead}}$ but is merely the space of quantum states. It
cannot have a physical analog because the physics is required
neither for $\bbH^{\sss\langle\rangle}$"=space nor for its `bare'
\lvs"=version~$\bbH$.

We call mathematical construction $\bbH^{\sss\langle\rangle}$ the
unitary space \cite{halm} or the Hilbert space, stipulating the
point that questions of topology and of infinities will be
considered in their own rights. Assuming that the question of the
(normed) topology on $\bbH$ is solved in the affirmative
(sect.~\ref{topology}), we arrive at the final result.
\begin{itemize}
\item \embf{The $\bo{3}$-rd theorem of quantum empiricism $($on
 Hilbert's space$)$}
 \begin{enumerate}
 \item[1.]The space of quantum states $\bbH^{\sss\langle\rangle}$ is an
  abstract vector space over $\bbC^*$, which has been equipped with
  the statistical"=length function \eqref{a2} and the concept of the
  quantum observable $\scr A$ as an orthogonal denumerable basis.
  All the models to this structure are isomorphic.

 \item[2.]The inner"=product superstructure \eqref{axiom} and
  normalization $\langle\bo\Psi|\bo\Psi\rangle=1$, not being a
  necessity for $\bbH^{\sss\langle\rangle}$, do formalize the
  general $\bbH^{\sss\langle\rangle}$"=calculus. Vector coordinates
  \eqref{=} in every $\scr A$"=base are calculated according the
  projection rule~\eqref{proj}; no use is required of the concept of
  operator.

 \item[3.]Statistical weights \eqref{BR} and \eqref{BR+} admit an
  invariant writing, which has the standard form of the Born rule
  \begin{equation}\label{BR++}
  \fr_k=\frac{|\frak a_k|^2}{|\frak a_1|^2+|\frak a_2|^2+\cdots}
  \quad\hence\quad \fr_{\bo\alpha}= \frac{\langle\bo\Psi|\bo\alpha\rangle
  \langle\bo\alpha|\bo\Psi\rangle}{\langle\bo\Psi|\bo\Psi\rangle
  \langle\bo\alpha|\bo\alpha\rangle}\;;
  \end{equation}
  or $\fr_{\bo\alpha}=|\langle\bo\Psi|\bo\alpha\rangle|^2$ under the
  normalization convention.

 \item[4.]Topology on $\bbH^{\sss\langle\rangle}$ is defined by function
  \eqref{a2} entailing the concept of a norm
  $\|\ketPsi\|^2\DEF\cal N[\ketPsi]$ and of metric
  $\varrho(\ketPsi,\ketPhi)=
  \|\ketPsi\mathbin{\hat{\smash-}}\ketPhi\|$.
 \end{enumerate}
\end{itemize}

\subsection{Comment on statistics and physical interpreting
the $\langle\bo\psi|\bo\varphi\rangle$}

The symmetry of the quantity $\fr_{\bo\alpha}$ with respect to
$\bo\Psi$ and $\bo\alpha$ allows us to forget that it was being
created for the $\scr A$"=bases as for collections of $N$ vectors
$\{\ket{\alpha_1}$, \ldots, $\ket{\alpha_{\sss N}}\}$. In this regard,
the very problem \eqref{F} implied, strictly speaking, not the pair
of data $\{\ketPsi,\ket{\alpha_j}\}$ but the whole $\scr A$"=basis.
But now, one may speak about the statistical weight of one abstract
state in the other:
\begin{equation}\label{star}
\fr= \brak[,]{\bo\Psi}{\bo\Phi}\brak[,]{\bo\Phi}{\bo\Psi}\qquad
\big(\text{symmetry $\ketPsi\rightleftarrows\ketPhi$}\big)\;.
\end{equation}
Thereby the observable number $\fr$ has acquired not merely an
invariant definition (no reference to the word basis) but has turned
into a \emph{binary} symmetrical structure on the entire
$\bbH^{\sss\langle\rangle}$. The structure ignores the orthogonal
remainders in linear expansions
\begin{equation}\label{ostatok}
\ketPsi=\frak c_1\bcdot\ket{\Phi_1}\+\cdots\;.
\end{equation}
There is no contradiction here, since each vector
$\ket{\Phi_1}\in\bbH$ may get to be an eigen one for a certain
$\scr A$"=instrument and each vector $\ketPsi\in\bbH$ is a carrier of
\emph{all} the statistics $\{\fr_1,\fr_2,\ldots\}$ under the arbitrary
$\ket{\Phi_1}$ and
$\{\ket{\Phi_2},\ldots\}\FED\ket{\Phi_1}^{\sss\bot}$.

Drawing an analogy to the classical statistics or to the probability
theory, a contrast is in place.
\begin{itemize}
\item \emph{The classical physics has no underlying linear
 theory}""---""theory of space $\bbH^{\sss\langle\rangle}$,
\end{itemize}
which is why such binarity is impossible. This point is known as a
problem with understanding/"!comprehending/"!defining and making sense
(`physicalization') of the quantum probability \cite{laloe, hren,
peres}. For example, it is obvious that for the purely
statistical/"!classical analog to the left/"!right hand part of
\eqref{ostatok}, \ie, for the data-set \ceils{$\fr$"=statistics}
\tplus\ \ceils{$\alpha$"=spectra} such as
\begin{equation*}
\big\{(\skew{1}\tilde\fr_1,\tilde\alpha_1),
(\skew{1}\tilde\fr_2,\tilde\alpha_2),\ldots\big\},\quad
\big\{(\fr_1,\alpha_1), (\fr_2,\alpha_2),\ldots\big\}\qquad
({}=\text{\Courier{StatData}})\;,
\end{equation*}
formula for calculating the $\fr_1$ by
`\Courier{D}$\widetilde{\text{\Courier{at}}}$\Courier{a}-vector'
$\{(\skew{1}\tilde\fr_1,\tilde\alpha_1),\ldots\}$ cannot exist. The
stat"=distributions $(\skew{1}\tilde\fr_1, \skew{1}\tilde\fr_2,\ldots)$
and $(\fr_1, \fr_2, \ldots)$ are not related in any way. There is no an
$\bbH^{\sss\langle\rangle}$"=theory analog in between.

In the physical theories, the aforesaid binarity makes it possible
to give meaning to such wording as \ceils{$\ket{{}\texttt{before}}$,
$\ket{{}\texttt{after}}$}, transitions
$\ket{{}\texttt{in}}\goto\ket{{}\texttt{out}}$, \etc. However these
mathematical equal-rights do not, as before, bear on the
reversibility $t_1\rightleftarrows t_2$ in time that is associated
usually with the `physical processes' $\ket{\Psi_{t_1}}
\mathrel{\rlap{\raise0.4ex\hbox{$\dashleftarrow$}}
\lower0.4ex\hbox{$\dashrightarrow$}} \ket{\Psi_{t_2}}$. The nature
of the quantum Hilbert space is free of (also the
physical/"!empirical) notion of time, because the abstract states are
\emph{not} tied to the chrono"=geometrical notions of
causality/"!(non)""locality/"!propagation/"!speed (of something, say of
light $c$) and to ``the objective determination of space"=time
phenomena'' (W.~Heisenberg, Nobel Lecture). Indeed, the
$\bbH^{\sss\langle\rangle}$"=theory does not yet have a collocation
``causality of the classical events''; none of these concepts have
been defined. Say, it is meaningless to bring the two states
$\ket{\Psi_{t_1}}$, $\ket{\Psi_{t_2}}$ into correlation with each
other in the context of causality/"!locality/"!determinism
\emph{without} conception `the observable'. But again, creation of
the latter has not yet been completed.

As remarked previously, unitarity $\skew0\widehat{\cal U}$ may be
weakened merely to a preserving the $\fr$"=structure \eqref{star} and
orthogonality: \ceils{unitarity $\tplus$ scalability}. Isomorphism
of the $\bbH^{\sss\langle\rangle}$"=space models remains under that
modification.

\section{On (a quantum) Pythagoras theorem%
\protect\footnotemark}\label{Pif}%
\footnotetext{Content of this section is an exhaustive
math"=explanation to the popular and illustrated
exposition~\cite{br4}.}

Having written the additive property \eqref{=2} in the
$\bbH_{\sss\scr A}$"=notation
\begin{equation}\label{add}
\cal N_{\!\!\!\sss\scr A}\big[a\bcdot\msf X \+
b\bcdot\msf Y\big]=\cal N_{\!\!\!\sss\scr A}\big[a\bcdot\msf X\big]+
\cal N_{\!\!\!\sss\scr A}\big[b\bcdot\msf Y\big]\qquad\forall\,
a,b\in\bbC\;,
\end{equation}
one can go further and adopt for the statistical length, by
etymology of this term and the quadratic form \eqref{a2}, the new
notation $\cal N_{\!\!\!\sss\scr A}[\msf X]\FED\|\msf X\|^2$. Then
\begin{equation}\label{abXY}
\|a\bcdot\msf X\+b\bcdot\msf Y\|^2=
\|a\bcdot\msf X\|^2+\|b\bcdot\msf Y\|^2\qquad \forall\, a,b\in\bbC\;,
\end{equation}
where the two in $\|{\cdot{\cdot}\cdot}\|^2$ does not yet mean the
squaring a number. Recalling now that the statistical content of
vectors implies their free scalability $\{\msf X\mapsto
a\bcdot\msf X\FED\itsf{x}\,, \msf Y\mapsto
b\bcdot\msf Y\FED\itsf{y}\}$ with preserving the orthogonality and
the observable meaning, we get
\begin{equation}\label{pif}
\|\itsf{x}\+\itsf{y}\|^2 = \|\itsf{x}\|^2+\|\itsf{y}\|^2\,;
\end{equation}
\ie, identity with the standard writing the Pythagoras theorem
$|\vec{\bo x}+\vec{\bo y}|^2=|\vec{\bo x}|^2+|\vec{\bo y}|^2$, in
which the first sign of `addition' should however be denoted by a
different plus. The passage \eqref{abXY} $\rightarrowtail$
\eqref{pif} may seem to be a formal `concealment' of quantum field
$\bbC$ into the $\bbR$"=reality of Pythagorean theorem only at first
glance. Therefore, let us get back to the concept of orthogonality
(sect.~\ref{sec-ort}).

\subsection{Quantum `inspection' of Pythagorean theorem}\label{Pif1}

The vocabulary that is involved in the theorem comprises the
following terms: triangle, side, direction, length, perpendicular,
addition, geometric square, angle, sum, the orthogonal, the right,
distance, area, diagonal, squaring, cathetus/"!hypotenuse, numerical
operations, \etc. Bearing in mind the fact that all of them appeal
to the familiar model on a plane, the list should be supplemented
with words about rotations, translation, and about reflections,
since handling the squares implies their geometric transporting;
squares are compared only `with the help' of group of motions.

In the natural/"!scholastic language, all these notions are considered
as \emph{real} entities that accompany triangles, squares, \thelike.
Though labeled by numerals, \emph{noon} of these entities are the
number itself. For instance, addition of the directed line segments
(forming a triangle), addition of areas (`square meters'), and
addition of usual numbers is far from being the same addition; not
to mention subtraction. See, \eg, commentaries and emphasis in
italics by Mordukha\u{\i}-Boltovsko\u{\i} concerning the
ancient"=Greek perception of the ``\emph{idea of a number}'' on p.~375
in \cite[Russian translation]{euclid} and notably the explanation to
the effect that the ``Euclidean $AB\Times AC$ is not in any way \ldots\
not in the sense as understood in arithmetic \ldots\ not
the~\emph{multiplicare}''; and also a comment on arithmetizing the
geometric images and ``quantities \emph{at all}'' in section
``\textbf{2.~Product of segments}'' on pages~297, 248, and 317. The ``area
is a quantity, the essence of which as a primary notion is not
defined by [Euclid]. \ldots\ he does not give a measuring the area by
\emph{number}'' \cite[pp.~286--7]{euclid}.

However, it is clear that Pythagorean theorem is a quantitative
statement about the aforementioned segments, which `visually add up'
to triangles with scalable sides. And this is what we call the model
of the vector space. Its operations $\{\+,\bcdot\}$, due to
difference between their nature and the arithmetical operations
$\{+,\times\}$, are not numerical but abstract in principle;
especially the unary (not binary) `multiplication' symbol~$\bcdot$.
Thus,
\begin{itemize}
\item the accurate (re)formulation of the theorem, one way or another,
 demands proceeding to the abstract vectors of the abstract \lvs
\end{itemize}
(the associating an affine space to this \lvs\ is not essential in
our context). As this takes place, no such quantities as
squares/"!\ldots/"!angles should be implied. These should be \emph{built
over} the \lvs, and relations between them should be ascertained.
Put another way, the `pure algebra of bare' \lvs\ is supplemented
with `calculus of the real entities'. But this is almost the same
situation that took place when deriving the Born rule \cite[thesis
$(\bullet)$]{br2}.

If, as usual for \lvs, we introduce the scalar product then the
theorem \eqref{pif} would boil down to the necessary and sufficient
condition of orthogonality of vectors
\begin{equation}\label{x+y}
\itsf{x}\mathrel\bot\itsf{y}\quad\hhence\quad
(\itsf{x}\+\itsf{y},\itsf{x}\+\itsf{y})=
(\itsf{x},\itsf{x})+(\itsf{y},\itsf{y})\qquad
(\text{field $\bbR$})\;.
\end{equation}
The theorem is thereby simplified; there remains a single identity
between the four entities $\{\bot, (\bcdot,\bcdot), \+, +\}$ without
invoking the vocabulary mentioned above. Thus formalization
discloses excessiveness of the usual terminology. The necessary and
the redundant elements are intertwined in meaning with each other,
however, the consideration does not end with this.

Orthogonality and scalar product are the derivative constructs of
the quadratic function \eqref{a2}. By tracking its emergence in
\cite{br2}, it is a simple matter to see that the derivation
procedure remains the same (and even simpler \cite[p.~3]{br4}) for
the field $\bbR$ as well, provided that the complex
$\smallc({*}\smallc)$"=involution is replaced with
$\frak a\mapsto(-\frak a)$. In this respect, the nature of the
scalar product and of the companion concepts""---angle and
orthogonality""---does not depend on the number field
(cf.~\cite[sect.~60]{halm}). Besides, the numerical
$\cal N$"=structure \eqref{a2} is so minimal construct \cite{br2}
that there is no need to introduce terminology of
lengths/"!\ldots/"!angles. Clearly, this is not about the quantum scheme of
things\footnote{Although the main quantum ideology is easily
visible: one requires a ``\emph{controlling the language over
itself}'' \cite[sect.~11.1]{br1}. In quantum situation, the natural
language creates the \lvs\ and then the language of physics. In
classical Pythagoras' geometry, the language of squares/"!\ldots/"!angles
in theorem itself \emph{follows from} the \lvs"=language; see below.}
since vocabulary of realities in the normal wording the theorem is
automatically recovered.

Meanwhile, in quantum course of action over linear manifold, the
issue of the intertwined terminology does not even arise. Logic of
\qt\ calls for severe separation of \lvs"=abstracta from observable
entities at the outset \cite{br1}. Thus,
\begin{itemize}
\item we arrive at the general conclusion about a single nature of the
 classical Pythagorean theorem and the Born rule \cite{br4}.
\end{itemize}

The analogies between Pythagorean theorem and Hilbertian sum of
squares were self"=apparent at all times; they are encountered
throughout the literature. But the issue lies in straightening out
the concepts, whereupon the rule ceases, as we have seen, to be a
postulate, and the physical `rationale behind' the \StatL\ can be
cast away. Consider now some details, without attaching much
distinction to Pythagorean and Bornian cases. All the more so the
origin of the square~$^2$ has already been ascertained and it was
ascertained in the quantum context.

The very first real quantity in theory is the \StatL; this is a
theoretical representative of the $\state\alpha$"=click number
\cite{br2}. Because the scalar"=product structure is fully
\emph{derivable} and orthogonality is independent of it
(sects.~\ref{sec-ort}--\ref{sc}) and derivable as well, we are
dealing with not anything else but `remaking' the theorem
\eqref{x+y} to definition""---""deducing and formalizing the
structural properties of the object \eqref{SL}
\cite[\textbf{Definition}]{br2}, from which formula \eqref{a2} emerges.
Therefore quantum statistics of Born \eqref{BR} is not provable
within an extended Hilbertian version of the classical Pythagoras
theorem \eqref{x+y} under generalizing \eqref{x+y} to the field
$\bbC$:
\begin{equation*}
\itsf{x}\mathrel\bot\itsf{y}\quad\hhence\quad
\big\lceil(\itsf{x}\+\itsf{y},\itsf{x}\+\itsf{y})=
(\itsf{x},\itsf{x})+(\itsf{y},\itsf{y})
=(\ri\bcdot\itsf{x}\+\itsf{y},
\ri\bcdot\itsf{x}\+\itsf{y})\big\rceil\qquad
(\text{field $\bbC$})
\end{equation*}
(cf.~exercise 4\,(c) in \cite[p.~123]{halm}). The Born formula was
being derived rather than being proved \cite{br2}.

Simplification of the aforesaid leads to fact that the quadratic
notation \eqref{a2} begins to be associated with the words
`geometry, Pythagoras, theorem', and symbol
$\|\itsf{x}\|^2\DEF(\itsf{x},\itsf{x})$ loses the square. With this
dropping, this symbol is subconsciously equipped with a meaning
being called the length of vector $\|{\cdot{\cdot}\cdot}\|$ and
having some properties in triangles. But remaking of theorem to
definition should not disappear. Notice that proofs of the
(direct/"!converse) Pythagoras theorem, wherein lengths do \emph{not}
appear, are well known. Such is the proof (complicated) and wording
of Euclid himself \cite[Proposition~47 and~48, pp.~46--48]{euclid}
in the language of adding the plain areas.

In point of fact, the aforementioned `different additions' say that
the classical formulation""---`the square of the (length of)
hypotenuse \ldots', by the quantum viewpoint, abounds with empirical
inaccuracies. It is utterly fundamental to claim: how and what's
being added, what is defined through what, whence the length, what
we have, what we have no, and what's being deduced. Following thus
the quantum spirit, we need to line up, as accurately as possible,
the strict hierarchy of `what from what', including the precise
indication `which addition', `which multiplication', and what is
understood by them at all.

\subsection{Additivity and scalability. What is length?}\label{dln}

The language intension of the \emph{additivity} of \StatL\
\eqref{Naj}--\eqref{=2} is a core point both in Born's rule and in
Pythagoras' theorem because the natural language is always necessary
\cite[sect.~3.1.1]{ludwig5}, \cite{salmon}; it is primary even for
foundations of mathematics \cite{frenkel, lakoff, chomsky},
\cite[Chs.~1, 4, 19, 21]{russel}. Being a concept that is
inseparable from $\ket\alpha$"=vectors of observable $\scr A$, the
additivity of \StatL\ gives birth to their orthogonality and, then,
to the scalar product. There appears the space
$\bbH^{\sss\langle\rangle}$ and, as a consequence of this
definition, the habitual language of the triangle sides should be
reorganized (see also sect.~\ref{qlang}) to the language of linearly
independent vectors of \lvs. Expressed differently,
\begin{itemize}
\item what the `side of triangle' is supplemented by under the term
 length/"!\ldots/"!square is defined as an additive function $\cal N$
 (property \eqref{add}), which is, upon action by number, expressed
 only through itself (non"=multiplied):
 \begin{equation}\label{mul}
 \cal N[\oper{\frak c}\,\ketPsi]=\msf C(\cal N[\ketPsi])\,,\qquad
 \cal N[\oper{\frak c}\,\ketPsi] \mathrel{\Over?=}
 \msf{const}\times\cal N[\ketPsi]\;.
 \end{equation}
\end{itemize}
That is to say, quantification $\cal N$ of real things
\cite[sects.~7.1--2 and Remark~16]{br1} is inseparable from the
operator nature $\ketPsi \mapsto
\oper{\frak c}\,\ketPsi\FED\frak c\bcdot\ketPsi$ of the abstract
number $\frak c$ \cite{kurosh}, \cite[\S\,7.2]{br1}.

The unit character of the quantity $\cal N$ under
creation\footnote{Metres, square metres, \thelike. This is also a
part of the natural"=language definition of $\cal N$, but it is
subject to `revision' through the rule \eqref{mul} too. In this
regard, the term \texttt{Stat\-Size} or \texttt{Stat\-Number} (statistical
number) would be better suited for \StatL.} calls for ascertainment
of its `multiplicativity' property \eqref{mul}, while the fact that
function $\msf C$ ought to become a multiplying does not follow from
anywhere. Planning to call $\cal N$, say, the length of vector/"!side
or to create the concept of a square/"!volume, we may not postulate in
advance the character of its scalability since, in \lvs, there has
been present an axiom that combines the action of
multiplication~$\bcdot$ and of addition~$\+$. This is the
distributive law
\begin{equation}\label{distr}
\frak c\bcdot(\ketPsi\+\ketPhi)=
\frak c\bcdot\ketPsi\+\frak c\bcdot\ketPhi\quad\hence\quad
\cal N\big[\frak c\bcdot(\ketPsi\+\ketPhi)\big]=
\cal N\big[\frak c\bcdot\ketPsi\+\frak c\bcdot\ketPhi\big]\,,
\end{equation}
and it is it that dictates what the rule \eqref{mul} has to be
\cite[p.~9]{br2}. This fact alone says that lengths, squares, and
volumes do exist not on their own account but are tied to the
abstraction \lvs, which is not so obvious when proceeding from their
everyday understanding or from the classical physics.

After having ascertained the quadraticity of the scaling, \ie\ once
eqs.~\eqref{mul}--\eqref{distr} together with involution
$\cal N[\frak c^*\bcdot\ketPsi]=\cal N[\frak c\bcdot\ketPsi]$ have
led to the expression
\begin{equation*}
\cal N[\frak c\bcdot\ketPsi]=|\frak c|^2\times \cal N[\ketPsi]
\end{equation*}
(see \cite[sect.~5]{br2} for details), one reveals a distinction in
operations $\{\bcdot, \times\}$ and disadvantage of the term
$\smallc({\bcdot}\smallc)$"=multiplication.
\begin{itemize}
\item The intuitive perception of words `to scale vector by a number'
 does \emph{not} furnish the naturally anticipated `to change its
 length'. The construct $\cal N$ gives rise, in case of field
 $\bbR$, the square on a vector and, in the quantum $\bbC$"=case, the
 \StatL. The question of length still stands.
\end{itemize}
Thus, leaving aside vectors and the \lvs\ itself, we conclude that
the square (generalized) rather than the length is a primary
quantitative entity both for the theorem and for the rule. But in
both the cases, there has been `hardwired for free' the concept of
orthogonality.

The reason of this phenomenon is of course the multidimensionality
of \lvs\ because, in case of $\dim\bbH=1$, the concept of linear
independence goes away, and additivity and length are trivialized
just into a number. Once $\dim\bbH>1$, there arise nonequivalent
bases and arbitrariness in coordinates, and intuitive `1"=dimensional
(\,$=$ quantitative) concept' the length must be created for vector.
This is a \emph{non}trivial action since vector""---even the
scholastic""---is an \emph{abstraction} far from merely the number.
The nature of the latter is roughly speaking the `number of
something' \cite[sect.~7.2]{br1}, while the `number of a
multidimensional' is an ill"=defined semantics. Revision of the
theorem is exactly what gives meaning to that.

As a result, the length ceases to be a 1"=dimensional structure
existing irrespective of the `2-dimensional concepts' of the right
angle, orthogonality, or square. Attention is drawn to the fact that
the concept of a square, nevertheless, does not arise as an object
on two vectors; it is not a binary construction. Speaking more
loosely, length is defined through a square root of something more
primary; cf.~definition
$d\ell=\sqrt{\smash{g_{\lambda\mu}}\,dx^\lambda dx^\mu}$ in geometry
or in gravitation theory. Therefore, when the term orthogonality is
dismissed in the classical case, there should disappear not only the
`subject of proving' in theorem \eqref{x+y} but also the length as a
notion. In doing so, vectors, their coordinates, and the visual
images"=segments still stand.

The ideology of non"=axiomaticity prohibits introducing the length
through a norm since it relies entirely on the above"=listed
intertwined terms. Indeed, what is the motivation for arising a
(new) concept of triangle
$\itsf{x}\mathbin{\hat\pm}\itsf{y}=\itsf{z}$ and whence the
associated familiar inequality \cite[p.~333]{fried},
\cite[p.~127]{halm}, had we not possessed the concept of a length?
The geometric (interpretative) intuition of the primary
lengths/"!norms is correct neither in \qt\ nor in the classical
geometry; more precisely, under the empirical arithmetizing these
theories.

The last structural property""---""invariance with respect to
involutions""---is obvious from the natural language. For example,
in the $\bbR$"=case, function $\cal N$ should not depend on changing
the direction in space $\vec{\bo x}\mapsto-\vec{\bo x}$. Indeed, a
line segment and its quantitative measure $\cal N$, being a
numerical add-on over vector, do not care the notion of an
origin/"!terminus inherent to geometric vector. There arises the
requirement $\cal N[(-1)\bcdot\itsf{x}]=\cal N[\itsf{x}]$.

The quantum constituent of the theory has already been elaborated at
length, therefore let us sum up with a focus on its `Pythagorean
part'.

\subsection{Theorem $\GOTO{}$ definition}

Let there be a problem, the model of which does in some way employ a
vector space. For us, this is the scholastic geometry and quantum
states. In view of autonomy of the \lvs"=axioms, any further theory
may only be built upon this \lvs\footnote{And perhaps the copies of
\lvs, when we create, say, the tensor products of vector spaces for
describing the multi"=particle quantum problems.} \cite[thesis
$(\bullet)$]{br2}. If the case in point is a quantitative theory""---and
this is our situation""---then these new number objects must be
supplemented with due regard to a numerical part of axioms of the
very \lvs\ \cite{fried, halm}. These numerical quantities have (must
have by the nature of the task) the language/"!semantic descriptions,
which are subject to mathematization. In the cases
considered""---""theory of Born and of Pythagoras, these
descriptions are formalized into the one minimalistic thesis
\cite[thesis $(\bullet\bullet)$]{br2}:
\begin{itemize}
\item[$\bullet\bullet$] Equip the linear independence \eqref{PsiE} with a numerical
 additive function $\cal N$ (of coordinate representatives of
 vectors), which is invariant under the number involutions.
\end{itemize}
Does the function exist and, if yes, are all bases (linear
independence) that allow this?

Among other things, this wording and the preceding rationale do in
fact `exorcize' the notion of a physical/"!illustrative/"!geometric
interpretation from the theory fundamentals, because the
\emph{structural properties of function $\cal N$, in and of
themselves, is what we really mean by the word interpretation}. See
Remark~4.1 in \cite{br2} for more detailed comments.

In the language of formulas, the proposed minimalism turns into the
rules, which we write down as applied to the plain Pythagorean case
($\dim V=2$).

An operator character of the notion of the number, \ie\ \eqref{mul},
has been implied at all times.
\begin{enumerate}
\item[\hypertarget{p0}{\red i)}] For the two scale"=related vectors
 $\tgreek{a}\mapsto \oper[1]c\,\tgreek{a}$, the quantities
 $\cal N[\tgreek{a}]$ and $\cal N[\oper[1]c\,\tgreek{a}]=
 \cal N[c\bcdot\tgreek{a}]$ must be related to each other:
 \begin{equation*}
 \cal N[c\bcdot\tgreek{a}]=\msf C(\cal N[\tgreek{a}])\;.
 \end{equation*}
\end{enumerate}
Then the two claimed properties follow.
\begin{enumerate}
\item[\hypertarget{p1}{\red ii)}] Additivity on the linearly independent
 vectors $\{\tgreek{a},\tgreek{b}\}$:
 \begin{equation}\label{Nab}
 \cal N\big[a\bcdot\tgreek{a} \+ b\bcdot\tgreek{b}\big]=
 \cal N[a\bcdot\tgreek{a}] + \cal N[b\bcdot\tgreek{b}]\qquad
 \forall\, a,\,b\in\bbR\;.
 \end{equation}
\item[\hypertarget{p2}{\red iii)}] Involutory symmetry:
 \begin{equation*}
 \cal N[-a\bcdot\tgreek{a}] = \cal N[a\bcdot\tgreek{a}]\qquad\forall\,
 a\,,\tgreek{a}\;.
 \end{equation*}
\end{enumerate}
As earlier, the words triangle/"!length/"!\ldots/"!angle are considered now
non"=existent. Certainly, the theory is meaningless without
invariance with respect to the changes of bases.
\begin{enumerate}
\item[\hypertarget{p3}{\red iv)}] Well"=definiteness (\,$=$~meaningfulness
 of the quantity $\cal N$):
 \begin{equation*}
 a\bcdot\tgreek{a} \+ b\bcdot\tgreek{b}=
 a'\bcdot\tgreek{a}' \+ b'\bcdot\tgreek{b}'\qquad\hence\qquad
 \cal N\big[a\bcdot\tgreek{a} \+ b\bcdot\tgreek{b}\big] =
 \cal N\big[a'\bcdot\tgreek{a}' \+ b'\bcdot\tgreek{b}'\big]\;.
\end{equation*}
\end{enumerate}

Though technically important \cite[pp.~12--13]{br2},
\cite[p.~4]{br4}, the latter point might well have been omitted.
Clearly, the introducing a function $\cal N$ on vectors
$\{\tgreek{a},\ldots\}$ is absurd if the $\cal N$ is incompatible with
the concept of equality $(\tgreek{a}=\tgreek{b}) \hence
(\cal N[\tgreek{a}] = \cal N[\tgreek{b}])$. One immediately reveals
that \emph{not each} linear independence \eqref{PsiE}, \ie, not all
of the $\{\tgreek{a},\tgreek{b}\}$"=\ and $\{\ket{\boe}\}$"=bases
admit the function $\cal N$ but only some special ones. The
construct does automatically produce what they have to be. All
possible bases of \lvs\ are separated into the orthogonal""---the
$\scr A$"=bases---and all the remaining abstract ones
(sect.~\ref{basis}).

As a result, the points \hyperlink{p0}{\red i)}--\hyperlink{p2}{\red
iii)} entail not only the Pythagorean square~$^2$ per se and
formalization \eqref{x+y} but also the angles and, literally, the
entire elementary geometry around the theorem. The attaching to this
math the vocabulary from sect.~\ref{Pif1} is a matter of harmonizing
the terminology with ordinary verbal vehicles. Similarly the
geometry of Born's space $\bbH^{\sss\langle\rangle}$.

Let us elucidate the aforesaid geometrically. We simulate the
relation $\tgreek{a}\mathbin{\hat\pm}\tgreek{b} =
\tgreek{g}\!_{\sss\pm}$ in the plane and \emph{assign} to this the
words triangle of a general position (generic). The triangle should
be thought of as abstract, without terms the lengths of sides and
(right) angle between them. In virtue of involution, those triangles
that admit the function
$\cal N[\tgreek{a}\mathbin{\hat\pm}\tgreek{b}] =
\cal N[\tgreek{a}]+\cal N[\tgreek{b}]$
(\,$=\cal N[\tgreek{g}\!_{\sss\pm}]$) should be pictured as having the
equal $\cal N[\tgreek{g}\!\!_\sp]$ and $\cal N[\tgreek{g}\!\!_\sm]$.
Consequently, these triangles ought to be parts of a quadrilateral
having equal diagonals. It is implied that the words diagonal and
quadrilateral have been defined through symbols~$\hat\pm$; we are
continuing to avoid the word `length'. Let us call that triangles
the \emph{rectangular} and label them by the standard symbol of the
right angle $\raise6.3pt\hbox{\rule{7pt}{0.7pt}} \hspace{-0.7pt}
\rule{0.7pt}{7pt}$. A most intelligible illustrations of this
material is given in \cite{br4}. This accomplishes the commentary
both to the abstract and to the `actual' `theorem' of Pythagoras,
which appears here as nothing more than definition
\hyperlink{p0}{\red i)}--\hyperlink{p2}{\red iii)}. Incidentally,
the $\cal N$"=calculus mathematics, as opposed to the familiar
Pythagorean extension of numbers $\sqrt{1^2+1^2}=\sqrt2$, is not
beyond the scope of the number"=rationality domain:
$\cal N[\tgreek{a}]+\cal N[\tgreek{b}]=1+1=\cal N[\tgreek{g}]=2$.

The situation has a parallel in topology when creating the concept
of `a line'. An arbitrary (continual) point set, being initially
considered as `merely a set' (`dispersed anywhere and ad~libitum',
the generic), ceases to be `arbitrarily dispersed', and we portrait
it as (\,$=$~it becomes) a line only after the set has been equipped
with an algebraic axiom of ordering $a<b$. By analogy, the nature of
the geometric term right angle is algebraic and lies solely in
existence and in equipping the \lvs\ with the (quadratic)
function~$\cal N$. It is this function that creates the right
angle---rather than the reverse, and it arises even prior to the
notion of an angle or of its numerical characteristic like
$\cos(\Over[0.0]{\ds\widehat{\quad}}{\tgreek{a}\,\,\tgreek{g}})$.

\subsubsection{Additivity, again}
\addcontentsline{toc}{subsection}{\qquad\qquad\sffamily\itshape Additivity,
again}

Now, the ideology of an additive function on linear independence is
a key to ascertain an analogy in the `brace' Pythagoras--Born.
\begin{itemize}
\item Having had in a single theory the two \emph{different}
 pluses---the abstract/"!multidimensional $\+$ and the
 number/"!one"=dimensional~$+$, we should declare a rule of their
 compatibility. This is what the formulas \eqref{Naj} and
 \eqref{Nab} do via function~$\cal N$; a number add-on over the
 \lvs"=abstraction \cite{br2, br4}. The `Pythagorean' and the
 `quantum' are not distinctive in this regard.
\end{itemize}
This fact, along with uncovering the meaning to the Pythagorean
theorem through the quantum theory, seems to be lacking in the
literature\footnote{Likely, had the origin of Pythagorean squares
\eqref{pif} through the additivity \eqref{Nab} been known, its
quantum $\bbC$"=counterpart \eqref{a2} would not be a postulate and
would long have appeared in the literature as a normal formula. In
consequence, the familiar Gleason theorem (and his ``frame
functions'') would become a self"=suggested corollary of this point
when involving the concepts of the mean and of the linear
operator.}. The geometry reduces to the pure algebra without
intertwining the terminology \cite[sects.~59--62]{halm}. The theorem
itself turns roughly speaking into a definition of the additivity,
and the Born quantum postulate""---into a theorem"=corollary of this
definition and into the structure of the Hilbert space.

This is the main conclusion one should draw for understanding the
theorem and the postulate, regardless of whether we are looking at
them in a quantum or in a classical manner, \ie, regardless of the
number field $\bbC$ or~$\bbR$. An oddity is that the
physical/"!quantum theory not only updated the standard formulation of
\lvs\ (see~Remark~15 in \cite{br1} and sect.~\ref{intro}) but also
`compelled us' to look more carefully at such an ancient theorem,
turning it into a definitio. Here, we do not touch upon the topic
about relationship of the theorem with the 5"~th Euclidean parallel
postulate""---the theorem depends on it---and with the familiar
Riemannian discourse on empirical bases of geometry. Notice that the
concept of the length of vector is still up in the air, there is no
need for it. The matter will remain the same when considering the
topology in~sect.~\ref{topology}.

\begin{comment}\label{two}
The $\cal N$"=additivity \eqref{Nab} resembles an analogous property
of the additive measure $\mu$ on a set \cite{ludwig3}:
\begin{equation*}
\mu(A\cup B)=\mu(A) + \mu(B)\quad \text{for } A\cap B=\varnothing\;.
\end{equation*}
However our $\cal N$"=object is being sought not as function on
(sub)sets/"!spaces of \lvs\ \cite{engesser} but on coordinate
developments \eqref{PsiE}. The numbers are produced form the other
ones: \ceils{abstract numbers of the field $\bbR$} $\rightarrowtail$
\ceils{quantities/"!magnitudes}. The more so as the function exists
not for all of the bases \eqref{PsiE}.
\end{comment}

\subsection{Quantum and classical language, revisited}

The emergence of further (mathematically unnecessary)
terms---""perpendicularity, length, angles, distances,
\etc---""reflects a property of the ordinary language `to simplify
itself' \cite{chomsky}, introducing the larger and derived concepts
to avoid the repeating and heaping the primary primitives and long
collocations.

\begin{comment} One might say that the turning a certain
(lengthy) verbal vehicle into the integrated whole, the `naming' it
a single term (say, square), or identifying `the self"=similars, the
likeness's, \ldots' are, in and of themselves, an act of abstracting,
which permanently has been present in thinking, giving birth to the
primitive elements of the math language: sets, families,
addition/"!union, quantities, abstract numbers, \etc\
\cite[sect.~7.2]{br1}.
\end{comment}

Therefore when working backwards \ceils{mathematics
$\rightarrowtail$ explanation}, there arise the word
`interpretation' and the problem of treatments in terms of
observable quantities. The nature of observables in \qt\ is the
known and long"=standing polemical matter \cite{laloe, dAriano}. This
is due to the fact that the natural language, having its free
reducing and reproducing the phrases, oftentimes does not conform to
requirement of coordination/"!consistency that is a must in theory. As
can be seen, even the classical Pythagoras theorem does, in a sense,
`acquire a problem' of interpretation, since its accurate
re"=enunciation changes the way of looking at the notion of the
length. The commentaries by Mordukha\u\i"=Boltovsko\u\i\ \emph{to the
language} of Euclid's Elements \cite[Russian translation]{euclid}
elucidate this point very well.

The natural language `perceives the length' as the first and evident
observable entity, while there is no place for it in the correct
statement of problem""---\ceils{the \lvs} $\tplus$
\ceils{\hyperlink{p0}{\red i)}--\hyperlink{p2}{\red iii)}}. The
object `the square' should be in its own right rather than a square
of a length. It can in no way be declared as the `two segments with
equal lengths and a right"=angle in between', although when
visualizing the geometry of \lvs\ this cannot be avoided because
this is a part of the interpretation language. But in quantum
Pythagoras' theorem""---the statistical"=length rule \eqref{a2}---the
situation is opposite and simpler. The sum of squares is the number
one observable, and the length of quantum vector is absent as a
notion. Here, the observables and the abstractions have been
severely distanced, and in doing so, it is futile to introduce the
former prior to the latter \cite[sect.~10.2]{br1}.

A more formal view of the situation delivers, strange as it may
seem, the most precise explanation. The point is that the quantum or
Pythagorean vectors $\{\ketPsi,\frak c\bcdot\ketPsi\}$, from the
vector-space `standpoint', are \emph{merely different} vectors, the
different elements of a set. The mathematical structure the \lvs\ is
not `aware of' our geometric ways for visualizing the number as an
operator: $\ketPsi\mapsto\oper{\frak c}\,\ketPsi =
\frak c\bcdot\ketPsi$. Hence, the picturing this operation as a co-
or non"=codirectional dilatation (over $\bbR$ and notably over
$\bbC$) does amount to a bringing the new `illegal'
words---""shortening/"!stretching/"!\ldots/"!rotation""---""into the theory
of \lvs\ as an abstraction and to an implicit interpreting and
introducing the concept of the length.

\subsubsection{Numbers and observables, again}\label{qlang}
\addcontentsline{toc}{subsection}{\qquad\qquad\sffamily\itshape Numbers and
observables, again}

As for the numbers, the situation is analogous. These, as operators,
are applicable both to abstractions and to anything just as we apply
the numbers to various units \cite[Remark~16]{br1}. But in theories,
they both are being \emph{created}.
\begin{itemize}
\item The essential difference in between the abstract
 $\{\bbR,\bbC\}$"=numbers and the quantitative $\bbR^{\sp}$"=entities
 should not be neglected.
\end{itemize}
Say, when handling the expansions \eqref{PsiE} and even \eqref{=},
we must not think/"!imagine they constitute something the treatable,
the real, or simultaneously existing in writings like
$0.6\bcdot\ket{{}\texttt{here}}\+0.8\bcdot\ket{{}\texttt{there}}$. The
quantum foundations do forbid the numerical explanations a~priori
\cite{ludwig5, silverman, br1}, and this has an analog, as we have
seen, in classical geometry. The language for that explanations may
be created only \emph{after} the numerical object \StatL/"!square, and
only after it the quantitative objects under creation may be
consistently `accompanied' by the physical adjectives: observable,
spectral readings, real, measurable, \etc\ (see sect.~\ref{6th}).

Even if we were to pursue the aforementioned goals, then the word
`constitute' should be implemented through a mechanism in its own
right---""introducing the concept of `the observable quantity' and
explanation as to why the linear operator (if any) comes about here.
As a consequence, to take an illustration, for the notorious
`problem of Schr\"odinger's cat'---""ignore for a moment its
meaningless\footnote{``When I hear of Schr\"odinger's cat, I reach
for my gun'', says S.~Hawking in interview by T.~Ferris (Pasadena,
California, 4~April 1983).}---the statement of `the problem' must be
corrected. The mechanism of the introducing ought to bring the basis
invariance (\,$=$~quantum noncommutativity) into play. That is, apart
from observable $\{\ket{{}\texttt{alive}}, \ket{{}\texttt{dead}}\}$
(`cat's momentum'), one requires at least one more ``pointer"=state''
set \cite{busch}; \eg, the `cat coordinate' $\{\ket{{}\texttt{left}},
\ket{{}\texttt{right}}\}$"=corner in the box. Formalization of these two
`observables' is the pt.~\hyperlink{p3}{\red iv)}.

\section{Topology on quantum states}\label{topology}

Non"=physicality and non"=mathematicity of prerequisites for the
theory (sect.~\ref{intro}) have an impact on the question of a
quantum"=state topology, which seems to be entirely a formal problem.
Why and how should we norm the quantum vectors \cite{ludwig3,
ludwig5}? To measure (?) the length (?) of quantum state?

\subsection{Numbers and open sets}

It is not correct to say that the two states physically differ
little from each other \cite{ludwig3} (or, \eg, the one approaches
the other), because in the natural language, the phrase `differ
little' implies the handling of observable entities, and the word
`little' means the `reified $\bbR$"=numbers'. However the states are
in no way producible and comparable physically \cite{fuchs}. This is
a task of quantum (meta)""mathematics, and it consists in
transporting the natural"=language notions of the `smallness,
approximation, smoothness', \etc\ from empirical language to the
arisen $\bbH^{\sss\langle\rangle}$"=abstracta. It is common knowledge
that this is implemented by topology \cite[Introduction]{burbaki}:
neighborhoods, closed/open sets, limit/boundary points, \etc.

The ideology of non"=axiomaticity does not allow us to employ the
familiar methods of turning the \lvs\ into a topological space by
metric or by an isomorphism between $\bbH$ and $\bbC^N$ with an
automatical importing the natural (product) topology $\bbC^N$
onto~$\bbH$. All this are the mathematical rather than empirical
ways to topologize the~$\bbH$. For the same reason we are not
concerned with topological equivalence of norms on \lvs, including
their compatibility with scalar product. The necessity of the very
concept of a norm for quantum states---in effect, the question of a
length in sect.~\ref{dln}---is the subject matter of the present
section.

Topology is needed not only for introducing the continuity or
continuing maps of vectors $\ketPsi$ into something, but also is a
necessity for the internal needs of the $\bbH$"=space itself:
continuity of algebraic operations $\{\+,\bcdot\}$ and the making
sense to infinite sums of abstractions \eqref{=} when
$\dim\bbH=\infty$. The latter point has been commented in
\cite[sect.~6b]{br2}---the topology must be determined by function
\eqref{a2}. By virtue of its uniqueness, it is `topologically'
necessary for $\dim\bbH<\infty$ too.

The reducibility to numbers""---the values of function
\eqref{a2}---is to be a subject of topology inasmuch as \emph{we
have no} criteria for a number"=free way of declaring the open-set
systems or neighborhoods in the quantum~$\bbH$"=abstraction.
Therefore, even if we were to introduce these objects, then the
$\ketPsi$"=states might enter the condition that specifies such sets
only through the only numeric characteristic the quantum vector has
possessed---its own statistical length
$\cal N[\ketPsi]\FED\|\bo\Psi\|^2$ (erasing the
$\ket{{}\texttt{ket}}$). Moreover, the $\|\bo\Psi\|$ is a real number,
which is why we get towards the $\bbR$"=topology. Some comments are
now in order.

Given what we have said about \StatL, the task of introducing a
topology should be associated with a task of the \emph{numerical}
convergence of a chain $(\ldots,S_n,S_{n+1},\ldots)$ of the partial sums
$S_n=|\frak a_1|^2+\cdots+|\frak a_n|^2$; complexness of numbers
$\frak a_k$ is of no significance here. That is to say, the question
of an \emph{abstract} topology per~se is converted for our $\bbH$
into the question about convergence of the real"=number sequence
$\{S_n\}_{n\to\infty}$. And this is always the subject of
mathematizing the phrase `differ little from' \cite{lakoff}. With
the natural understanding of the number, this phrase is formalized
into the familiar inequality $|S_{n+1}-S_n|<\eps$. In turn, such a
difference $|b-a|$ can be reformulated without usage of
arithmetic""---the subtraction operation""---but only with using the
natural ordering~$<$:
\begin{equation}\label{top}
\begin{array}{c}
|b-a|<\eps\quad\hence\quad a<x<b\\[1ex]
\text{\ceils{$\eps$"=neighborhood,\quad$x$ is an element of
the open set $(a,b)$}}
\end{array}\;.
\end{equation}

It may be added that the natural ordering $<$ had entered a
definition of $\bbR$"=numbers when they were as yet arising in
quantum theory; through the ensemble"=accumulation procedure
\cite[sects.~2.5, 7.3]{br1}. If needed, that ordinal and
quantitative definition may be formalized into the set"=theoretic
inclusion~$\subset$. Then such notions as `little/"!nearly/"!\ldots/"!almost'
are represented mathematically by the `small $\eps$"=quantities
(cardinalities of sets) that are contained in between' numbers $a$
and $b$; the writing $a\subset\cdots\subset b$. Without such an
intension of the 1"=dimensional intervals \eqref{top} and of their
lengths $|{\cdot}{\cdot}{\cdot}|$, even the natural language has
been blurring.

\begin{comment}\hypertarget{p4}{}
More formally, one may draw on the following point, of which the
proof is omitted. Having had a (well"=defined, single"=valued)
function/"!map $\cal N$ from (as yet) non"=topological $\bbH$ onto the
numbers $\bbR$ (with natural topology), the preimage $\cal N^{\sm1}$
of open intervals on $\bbR$ delivers automatically a family of open
sets in $\bbH$. Clearly, the $\cal N$ is a function to be naturally
identified with the quantum one \eqref{a2}, and precisely a function
map into the numeric domain, being the only function that has been
motivated empirically. In its turn, the uniqueness of the natural
topology on $\bbR$ is also known, as each open set on the real line
is a countable union of intervals \eqref{top}. If a given $\bbH$ has
been equipped, as in our case, with algebra like $\{\+,\bcdot, +,
\times\}$, then the checking it for continuity is a math problem in
its own right. This should also include the supplementary questions
of the topology axiomatics""---the countability/"!separation
$T$"=axioms \cite[sect.~I.8]{burbaki}, which are essential for the
numerical convergence and for the meaningful concept of a limit
\cite[sect.~I.7.3]{burbaki}. Notice that for $\bbR$"=numbers with the
natural topology, all these axioms are met.
\end{comment}

Now, let us adopt that the notions `little difference, continuity',
\thelike, empirical as they are, have to be introduced and applied
to the $\ketPsi$"=abstracta. That is called for by `rigorising'
\cite{kronz} the $\bbH$"=calculus' constructed above: topological
completeness of the $\bbH$"=space, linear operator as a `smooth' map
$\bbH\mapsto\bbH$, its matrix elements, \etc. When reasoning about
continuity, the technically precise and universal term the abstract
open-set system is then substituted for the `small
quantities/"!numbers', \ie, for \emph{the} $\eps$"=neighborhoods
\eqref{top}.

\subsection{Are the norm and metric necessary?}

Thus the two possibilities are available. The first one---a
commonplace in physics""---is the axiomatical introducing the
concept of a length/"!norm $\|\bo\Psi\|$ and postulating its relation
to Born's square~$|{\cdot}{\cdot}{\cdot}|^2$, since each sum of
squares is a certain square:
\begin{equation*}
\ketPsi=\frak a_1\bcdot\ket{\alpha_1}\+
\frak a_2\bcdot\ket{\alpha_2}\+\cdots=\frak b\bcdot\ket\beta
\quad\hence\quad
|\frak a_1|^2+|\frak a_2|^2+\cdots=|\frak b|^2=\|\bo\Psi\|^2\;.
\end{equation*}
The $\bbH^{\sss\langle\rangle}$"=space mathematics is accomplished
then by the scheme:
\begin{equation*}
\text{norm $\|\bo\Psi\|$} \quad\rightarrowtail\quad \text{metric
$\rho(\ketPsi,\ketPhi) =
\|[1]\![4]\ketPsi\mathbin{\hat{\smash-}}\ketPhi\|[1]\![4]$}
\quad\rightarrowtail\quad \text{$\eps$"=topology \eqref{top}}\;.
\end{equation*}
The second option is to justify step-by-step a way of \emph{regular
deducing the axioms} of norm, \ie, to `take the square root' of the
\StatL\ \eqref{a2} and respect the algebra $\{\+,\bcdot\}$.

It is not improbable that despite the fundamental
multi"=dimensionality of the $\bbH$"=mathematics, one can even make do
without superstructure $\|\bo\Psi\|$ and, as the $\bbR$"=numbers
guide us (\hyperlink{p4}{{\blue Remark~4}}), reduce topological
examination for the $\{\ketPsi, \frak a, \+, \bcdot\}$"=algebra
solely to the 1"=dimensional language $\eps$-$\delta$. Speaking a
little simplistically, the very concepts of an abstract open set and
of norm can turn out to be an artifact for the quantum"=state
topology. This, of course, is not to say that it is not worth
adopting the norm, terminologically, as a usable concept.

We are inclined to believe that one can get around the first option
in quantum mathematics and reduce the second. To all appearances,
\emph{all the} ingredients of Hilbertian abstracta we have
considered in the present work may be viewed not as the postulated
structures but rather as the deducible ones. It is generally
tempting to infer that the norm $\ell^2$"=topology is, in a sense,
natural and inevitable. At least, it does really stand out from the
others, in particular, from~$\ds\ell^\msf{p}$.

Let us now return to \hyperlink{p4}{{\blue Remark~4}}, \ie\ to
reservations about the countability axioms, about the topological
indistinguishability of norms on \lvs, and to the fact that the
total number of axioms that pertain to the topology on $\bbH$ is
likely more than a dozen\footnote{Including, \eg, a numerical axiom
of Archimedes \cite{burbaki}. We mention this example, inasmuch as
the dismissal of quite low"=level axioms is known not only in
mathematics""---non"=Archimedean fields---but also in the $p$"=adic
redeveloping the quantum theory itself \cite{hren0}.}
\cite[Ch.~I]{burbaki}. All of them are essentially abstract and far
from empiricism, much less from physics. These will need to be
(re)considered, but this is a \emph{solely mathematical} problem.

In the physical literature, the issues of this sort are often just
ignored \cite{muller, aaronson+, silverman}, and quite clear why;
see, however \cite{wightman}. It is obvious that the concern, in its
full generality, will move deeply into the domain of mathematical
logic or even the foundations of mathematics \cite{frenkel, lakoff,
russel} rather than of (math)physics; not to mention the quantum
foundations. One must have stopped somewhere because the questions
will simply lose their significance for these very foundations.
Indeed, should discussions on the \qt"=elements take up the
separation $T$"=axioms (Hausdorff, Kolmogorov, \ldots), the Urysohn
theorems on metrizability/"!immersion, or the countable base of
topology, \etc? All these are the things that ensure separability of
$\bbH$, uniqueness of the limit, and, in general, meaningfulness of
the math we have been accustomed.

Let us summarize briefly the stuff above. The origin of an abstract
structure of the Hilbert \qt"=space, as well as of a Born-rule fromat
\eqref{BR++}, may be thought of as the resolved question. The
sought-for procedure is described, as was stated in the previous
study \cite{br2}, by the axiom-free continuation of the axiom-free
scheme~\eqref{scheme}:
\begin{center}
\ceils{micro-events' accumulation}\quad$\hence$\quad\ceils{\lvs\quad\tplus\quad
\eqref{a2}}\quad$\hence$\quad \ceils{Hilbert space
$\bbH^{\sss\langle\rangle}$}\;.
\end{center}

\section{Concluding remarks (physics)}\label{6th}

Non"=axiomaticity and non"=physicality of the
$\bbH^{\sss\langle\rangle}$"=construct and the `observable' \StatL\
lead us to the general conclusion in the context of Hilbert's sixth
problem \cite{accardi+, corry}.

On the one hand, the problem calls for `quantization'; on the other,
recalling von~Neumann's programme \cite{neumann}, the
straightforward applying and adhering to the ideology \ceils{axioms
$\rightarrowtail\cdots \rightarrowtail$ interpretations} is not
possible. The statement of the problem cannot disregard the language
semantics because semantical circularities are very well known
\cite{laloe, silverman}, and informal semantic analyses are needed
prior to any formal reconstruction \cite{salmon, chomsky}.
Semantics, in turn, begins with empiricism of quantum micro"=events
\cite[sect.~2]{br1}. More to the point, we have seen from
sects.~\ref{basis}, \ref{sc} and \ref{Axiom} that one may dismiss
not only the bulk of mathematical axiomatics but also the physical
aspects: quantum measurements \cite[p.~xiii; ``there can be no
quantum measurement theory'']{peres}, quantum probability, quantum
transitions/"!leaps, dynamics of observables, interaction, preferable
bases, \etc. Instead of the habitual `backbone' \ceils{math} \tplus\
\ceils{physical principles}, axiomatics of physics is replaced by a
single underlying construct free from the words ``axiomatische
Methode/"!Behandlung'' (Hilbert) \cite{corry}.

The situation is close to that of interpreting the formal languages
in mathematical logic \cite{frenkel, ludwig5}. Therefore the Hilbert
problem is not solvable without streamlining the nomenclature and
without hierarchy and splitting the language in use into the
\begin{itemize}
\item[\hypertarget{1+}{$\Under[0]{\smash{\lower3.5ex
 \hbox{\large$\DOWN$}}\phantom{^\bullet}} {\red1^\bullet}$}]

 Meta"=language: micro"=events, ensembles thereof, theoretical
 primitive $\statePsi\GOTO{\sss\scr A}\state\alpha$ \cite{br1},
 setting the macro"=environment by the conception `the same'
 \cite[sect.~5.4]{br1}, operatorial/"!quantitative/"!ordinal meaning to
 the number, \ldots

\item[\hypertarget{2+}{$\Under[0]{\smash{\lower3.5ex
 \hbox{\large$\DOWN$}}\phantom{^\bullet}} {\red2^\bullet}$}]

 Object language: $\{\bb R,\bbC\}$"=numbers, $\bbH$- and
 $\bbH^{\sss\langle\rangle}$"=structures (Hilbert space), spectra,
 \ceils{mixture of orthogonal states
 $\{\ket{\alpha_1}^{{\sss(}\varrho_1^{}\sss)}$,
 $\ket{\alpha_2}^{{\sss(}\varrho_2^{}\sss)},$ \ldots\}}
 $\rightarrowtail$ \ceils{statistical operator $\oper\varrho\,$},
 quantum statistics, \emph{the concept of a mean}, linear operators,
 self-\hskip0pt adjointness, the notion of the quantum system with
 an inner composition (multiparticleness and tensor products of
 $\bbH^{\sss\langle\rangle}$"=spaces), \ldots

\item[\hypertarget{3+}{$\Under[0]{\smash{\lower3.5ex
 \hbox{\large$\DOWN$}}\phantom{^\bullet}} {\red3^\bullet}$}]

 Math"=physical theories \cite{ludwig5}: instrumental readings,
 numerical measurement, spacetime continuum, quantum nature of the
 metric tensor $g_{\alpha\beta}\, dx^\alpha dx^\beta$,
 (non)""observables, causality, locality, dynamical
 equations/"!variables/"!fields/"!potentials, symmetries and
 $\oper{\cal U}$"=operators, the concepts of an interaction and a
 closed system, the ideology of quantizing a model, of the
 Hamiltonian, of the Lagrangian, of action, \ldots

\item[\hypertarget{4+}{$\red4^\bullet$}] Language of (physical)
 interpretations: tangible bodies, masses, forces, waves, observable
 phenomena, analogies in between, their descriptions through each
 other, the explanation language, \ldots.
\end{itemize}
In this, each language is created from the previous ones through the
natural language. The interpretation problems, paradoxes of \qt, and
questions of reality \cite{fuchs}, \cite[Introduction]{hren},
\cite[p.~10]{peres} disappear in the sense that they become a task
of the axiom"=free \emph{creating} the languages of the
(math)physical reasoning \hyperlink{3+}{$\red3^\bullet$} and
\hyperlink{4+}{$\red4^\bullet$}. Their terminology""---this we stress with
emphasis""---is forbidden in languages \hyperlink{1+}{$\red1^\bullet$} and
\hyperlink{2+}{$\red2^\bullet$}. Understandably, the very meaning of the
words `explaining, the explanation language' does always imply a
hierarchy of vocabularies in use \cite{ludwig5, chomsky}. The main
difficulty here is that the separation \ceils{pre-math, math,
pre-phys, physics} in transition \ceils{\hyperlink{1+}{$\red1^\bullet$}
$\goto\cdots\goto$ \hyperlink{4+}{$\red4^\bullet$}} breaks the habit of
ratiocinating in the seemingly inevitable language of realistic
notions and of human intuition.

All this much having been said, ``the formal frame for quantum
theory'' (von~Neumann) does not seem to require axiomatics in the
ordinary sense of the word, if we do not postulate initially the
concepts like metric and the space"=time as a topological continuum
with a dimension $D+1$ ($D=3$?). In particularly, the turning the
Hilbertian unitarity $\skew0\widehat{\cal U}$ (sect.~\ref{Axiom})
into the unitary dynamics $\skew0\widehat{\cal U}(t)$ (and
introducing a Hamiltonian) is a quite nontrivial act that
\emph{should be motivated} in the same manner as the initial
emergence of unitarity itself \cite[sect.~5]{br2}. We are speaking
here of quantum mechanics, although this notion should also be
(re)created. It is not difficult to foresee that the completion of
the language~\hyperlink{2+}{$\red2^\bullet$} is a more or less technical
task; not addressed in the present work (4"~th theorem).

To sum up, the coherent strategy for constructing the physics of
\qt-fundamentals consists in setting up the
languages~\hyperlink{3+}{$\red3^\bullet$} and \hyperlink{4+}{$\red4^\bullet$}
and should not recourse to the classical conceptions. In other
words, this \emph{needs not be a relativistic \qft"=generalization of
quantum mechanics} followed by a quantizing the gravity; to be
renormalizable? by quantizing the fields? Instead, the strategy must
constitute an ab~initio direct creation of a framework for an
entirely (Poincar\'e/"!generally) covariant theory, within which the
familiar ingredients""---the concepts of the gauge/"!observable fields
and of a particle, equations of motion, unitary/"!Hermitian operators,
Wightman's axioms \cite{wightman}, representations of the (local)
invariance groups---are being created on a regular basis of the
abstract Hilbert $\bbH^{\sss\langle\rangle}$"=space and a brace
thereof with the $(\bo x,t)$"=continuum. By this we mean the
following premise.
\begin{itemize}
\item \embf{The quantum origin of coordinates and fields}
 \begin{itemize}
 \item [] The classical and the quantum \emph{continual} objects---the
  spacetime coordinates $(\bo x,t)$ and fields $\{u(\bo x,t)$,
  $\bo A(\bo x,t)$, \ldots, $\skew1\widehat\psi(\bo x,t)\}$---may come
  to theory only from continuality of the quantum"=base changes
  $\big\{\ket\alpha \mathrel{\Over{\sss\GL[{}]}{\goto[3]}}
  \ket{\alpha'}$, $\frak a \mathrel{\Over{\sss\GL[{}]}{\goto[3]}}
  \frak a'\big\}$. From this, there also springs the (Einstein's
  idea of) `liberation' from coordinates: the follow-up concept of
  covariance of the dynamical field equations (the relativity
  principle), the concept of the vector/"!tensor/"!spinor
  representations of the specific (Poincar\'e, Lorentz, \ldots) groups,
  \thelike.
 \end{itemize}
\end{itemize}
See also the endings of sects.~8.3 and 9.3 in \cite{br1}. These
questions will be addressed elsewhere.

\end{document}